\documentclass[12pt]{iopart}
\usepackage{iopams,graphicx}
\def\H{{\mathcal H}}
\def\O{{\mathcal O}}
\def\<{\langle}
\def\>{\rangle}
\def\a{\alpha}
\def\b{\beta}
\def\d{\delta}
\def\e{\varepsilon}
\def\w{\omega}
\def\kT{{k_{\rm B}T}}

\begin{document}

\title
[Theory of electron-phonon renormalization and phonon-assisted optical absorption]
{Unified theory of electron-phonon renormalization and phonon-assisted optical absorption}

\author{Christopher E. Patrick and Feliciano Giustino}
\address{Department of Materials, University of Oxford, Parks Road, Oxford, UK, OX1 3PH}
\ead{feliciano.giustino@materials.ox.ac.uk}

\begin{abstract}
We present a theory of electronic excitation energies and optical absorption spectra which 
incorporates energy-level renormalization and phonon-assisted optical absorption within
a unified framework. Using  time-independent perturbation theory we show how
the standard approaches for studying vibronic effects in molecules and those for addressing
electron-phonon interactions in solids correspond to slightly different choices for the 
non-interacting Hamiltonian. Our present approach naturally leads to the Allen-Heine 
theory of temperature-dependent energy levels, the Franck-Condon principle, the Herzberg-Teller 
effect, and to phonon-assisted optical absorption in indirect band gap materials.
In addition our theory predicts sub-gap phonon-assisted optical absorption in direct gap
materials, as well as an exponential edge which we tentatively assign to the Urbach tail.
We also consider a semiclassical approach to the calculation of optical absorption spectra 
which simultaneously captures energy-level renormalization and phonon-assisted transitions 
and is especially suited to first-principles electronic structure calculations.
We demonstrate this approach by calculating the phonon-assisted optical absorption
spectrum of bulk silicon.
\end{abstract}

\submitto{\JPCM}
\maketitle

\section{Introduction}

The development of electronic structure methods for studying many-electron systems in solid-state
physics, nanoscience, and materials science, represents a major success of computational 
condensed matter research during the past three decades~\cite{Martin_book,Jones1989,Louie_book}. As the efficiency, 
availability, and accuracy of electronic structure techniques improve, the study of electron-phonon 
interactions from first principles is also becoming more accessible, and is drawing increasing interest 
among researchers. 

The electron-nuclear interaction is ubiquitous in the photophysics of solids. For example this
interaction renormalizes the excitation energies~\cite{Allen1976}, modifies the strength of optical 
transitions through phonon-assisted processes~\cite{Bassani_book}, and underpins intriguing phenomena 
such as the exponential Urbach tail~\cite{Urbach1953}. Yet despite the fundamental theory of indirect 
absorption being laid down more than half a century ago~\cite{Cheeseman1952,Hall1954,Fan1956}, it was not 
until very recently that a first-principles calculation of phonon-assisted absorption appeared in the 
literature~\cite{Noffsinger2012}. It took a similarly long period of time to combine first-principles 
techniques with the semiclassical theory of the temperature renormalization of band 
structures~\cite{Allen1976,Giustino20102,Cannuccia2011,Marini2008,Gonze2011}.

It is interesting to contrast these relatively rare solid-state calculations with the field of computational 
quantum chemistry. Here, established methods for calculating optical properties including the quantum 
motion of nuclei are implemented in widely-used software packages~\cite{Santoro2008}. 
Due to the different length scales addressed in solid-state physics and quantum chemistry,
the study of electron-nuclear interactions in these two areas
has followed distinct evolutionary paths 
leading to the development of different terminology. 
In quantum chemistry we find references to potential energy surfaces,
excited-state nuclear wavefunctions, and Franck-Condon factors~\cite{Atkinsbook}, while in solid-state
physics the nuclear motion is usually described in terms of a bosonic phonon field~\cite{Inksonbook}.
Yet since a truly first-principles approach should maintain its validity across the length 
scales, there should be no fundamental barrier to obtaining a unified description of the effects of 
quantum nuclear motion in molecules and solids.

Motivated by these simple considerations we set out to develop a unified conceptual framework for describing 
electronic excitations and optical absorption spectra. Ideally this new framework should encompass
and connect the current methods used for studying small molecules and extended solids.
In order to keep the presentation as general as possible, we start from the 
many-body Schr\"odinger equation for electrons and nuclei, and make no assumptions on the
practical approximations required to address the electron many-body problem. 

In order to describe 
electrons and nuclei on an equal footing we use {\it time-independent} perturbation 
theory~\cite{Sakuraibook}. This choice, which is at variance with standard approaches to the 
electron-phonon interaction in solids~\cite{Bassani_book}, is important in order to develop a unified 
framework. In addition this choice is useful for clarifying interesting aspects of the electron-phonon 
physics in solids which have been missed in the recent literature.

The choice of the non-interacting Hamiltonian in the perturbative expansion is not unique. We show
how two distinct choices lead naturally to the ``molecular picture'' or to the ``solid-state picture''. 
Having identified the non-interacting Hamiltonians we then
obtain correction terms for the electron-nuclear states and energies in powers of the electron-phonon 
coupling in each case. 

We then employ this power expansion to describe electronic transitions,
specialising to the case of optical excitations in the visible/UV range
(as opposed to e.g.\ vibrational infrared spectroscopy).
This analysis allows us to establish that phonon-assisted indirect 
absorption in solids and the Herzberg-Teller effect in molecules stem from the same perturbative term. 
As a byproduct of our analysis we obtain the temperature renormalization of excitation energies 
in solids, sub-gap absorption both in direct and indirect-gap materials, and an exponential absorption
edge similar to the Urbach tail. We also discuss a promising semiclassical approach for calculating 
phonon-assisted and vibron-assisted optical absorption in solids and molecules.

The manuscript is organised as follows. In section~\ref{sec.basicintro} we introduce the key 
quantities needed for describing the interaction between electrons and nuclei in molecules and solids. 
In section~\ref{sec.heuristic} we provide a heuristic discussion of how to include quantum nuclear 
effects in electronic structure calculations. This section allows us to review the concept of 
phonon-induced renormalization of band structures. In sections \ref{sec.pert}--\ref{sec.urbach} 
we tackle phonon-assisted absorption and energy-level renormalization from the quantum mechanical 
viewpoint within perturbation theory. We investigate both molecules (section~\ref{sec.mol}) 
and solids (sections \ref{sec.solids}--\ref{sec.urbach}). In section~\ref{sec.SC} we bring 
together the molecular picture and the solid-state picture and discuss a method of calculating optical 
absorption spectra including quantum nuclear effects across the length scales.
A practical demonstration of the method is provided in section~\ref{sec.Si}, where 
we calculate the energy-level renormalization and optical absorption spectrum of 
bulk silicon.
In section~\ref{sec.indir_sol_littleg} we discuss briefly some technical aspects
concerning the electronic Hamiltonian.
Finally
in section~\ref{sec.conclusions} 
we summarize our main findings and offer our conclusions. In this section we also highlight 
the key results and equations of our work. 

\section{The joint electron-nuclear system: notation and key approximations}\label{sec.basicintro}

\subsection{The electron many-body problem}\label{sec.basicelec}

We consider a system consisting of electrons and nuclei, and we use $r$ and $R$ to denote
the entire sets of electronic coordinates ${\bf r}_1, {\bf r}_2,\dots,{\bf r}_N$ and nuclear 
coordinates ${\bf R}_1,{\bf R}_2,\dots,{\bf R}_M$, respectively. The Hamiltonian of this system 
is given by:
  \begin{equation}\label{eq.fullH}
  \H = H_e^{R}(r) + T^R + W^R,
  \end{equation}
with $T^R$ the nuclear kinetic energy operator, $W^R$ the nucleus-nucleus Coulomb repulsion,
and the electronic Hamiltonian $H_e^{R}$ given by:
  \begin{equation}\label{eq.elecH}
  H_e^{R}(r) = T_e(r) + W_e(r) +  V_{en}^R(r).
  \end{equation}
In this expression $T_e(r)$ is the electronic kinetic energy operator, $W_e(r)$ the electron-electron
Coulomb repulsion, and $V_{en}^R(r)$ is the attractive Coulomb interaction between electrons and nuclei.
For each set of nuclear coordinates $R$ the formal solutions of the time-independent Schr\"odinger 
equation associated with the Hamiltonian $H_e^{R}(r)$ are obtained by solving:
  \begin{equation}\label{eq.He}
  H_e^R | \Psi^R_\a \> = E^R_\a | \Psi^R_\a \>,
  \end{equation}
where $\a$ may refer to a discrete index or to a continuous variable. By construction the eigenstates
$\Psi^R_\a (r)$ define a complete basis, ${\sum}_\a | \Psi^R_\a \> \< \Psi^R_\a| = 1$, where the sum 
should be understood as an integral in the case of continuous indices. Starting from the eigenenergies 
$E^R_\a$ we define the potential energy surface (PES) $U^R_\a$ in the electronic state $|\Psi^R_\a\>$, 
so as to make contact with the standard quantum chemistry literature:
  \begin{equation}\label{eq.PESe}
  U^R_\a = E^R_\a + W^R.
  \end{equation}
This quantity essentially describes the potential energy landscape seen by the nuclei when
the electronic and nuclear subsystems are considered as completely decoupled, with the electrons 
occupying the quantum state $|\Psi^R_\a\>$. At this stage, (\ref{eq.PESe}) should be regarded merely 
as a formal definition. In anticipation of the following discussion of electronic excitations and optical 
transitions we also introduce the electron excitation energy $\e_\a^R$ corresponding to the 
Hamiltonian in (\ref{eq.He}):
  \begin{equation}\label{eq.elec_excite}
  \e^R_\a = E_\a^R - E_0^R,
  \end{equation}
where the subscript ``0'' labels the electronic ground state, and we are considering neutral excitations. 
For notational convenience, in the following we will indicate the many-body electron eigenstates 
and the excitation energies obtained from (\ref{eq.He}) at the equilibrium nuclear coordinates $R = R_0$ 
as $\Psi_\a (r)$ and $\e_\a$, respectively (i.e.\ without explicitly 
indicating the superscript $R_0$).

Throughout the manuscript we will assume that the many-body electron wavefunctions
$\Psi^R_\a (r)$ are already known or can be calculated within some reasonable approximation.
In the simplest approach one could start from density-functional theory (DFT)~\cite{Hohenberg1964} 
and represent both the electronic ground-state and excited states using Slater determinants 
of Kohn-Sham wavefunctions~\cite{Kohn1965}. In this case the optical transitions would be
described within the independent-particle approximation, and the excitation energies would 
correspond to differences between the Kohn-Sham eigenvalues of empty and occupied single-particle states.
Alternatively one could use more advanced methods specifically designed 
to describe neutral excitations, such as time-dependent DFT~\cite{Runge1984}, or the Bethe-Salpeter 
approach~\cite{Onida2002,Onida1995,Rohlfing1998}. Similarly, in the case of quantum chemistry 
calculations there exist many options for calculating ground and excited states, from the coupled 
cluster method to multireference configuration interaction~\cite{Shavittbook}. Time-dependent DFT 
is also a very popular choice in this area, mostly within the standard Casida formulation~\cite{Casida1995}. 
The conclusions of our work do not depend on the specific approximations made in order to solve 
the electron many-body problem. The only requirement is that the chosen methodology be capable 
of yielding a reasonable optical absorption spectrum at 
fixed nuclei.

\subsection{Nuclear dynamics}

In order to address nuclear dynamics it is convenient to start from the potential energy surfaces
defined by (\ref{eq.PESe}). Using these surfaces we can introduce one nuclear Schr\"odinger equation 
for each electronic state $|\Psi_\a^R\>$:
  \begin{equation}\label{eq.molnucH}
  \left(T^R + U^R_\a \right)| \chi_{\a n}\> = E_{\a n}| \chi_{\a n}\>.
  \end{equation}
Here the nuclear wavefunctions $\chi_{\a n}(R)$ carry both the index $\alpha$ which specifies the PES,
and the index $n$ labelling the solutions of (\ref{eq.molnucH}). It is important to keep in mind that 
(\ref{eq.molnucH}) is merely a definition, and serves only as a starting point for the 
following analysis.

Despite its apparent simplicity, (\ref{eq.molnucH}) conceals an important subtlety. The definition of the 
potential energy surface $U^R_\a$ through (\ref{eq.PESe}) requires that the label $\alpha$ 
can be used to uniquely identify one electronic state for different nuclear configurations $R$. This is only possible 
when the states are non-degenerate for all values of $R$. The crossing of PES at certain nuclear
configurations, which are referred to as conical intersections in the quantum chemistry literature~\cite{Domcke2012}, 
require a separate discussion and will not be addressed here (see \ref{app.degen}). 

For future reference, in the case of the electronic ground state  
(\ref{eq.molnucH}) trivially becomes
  \begin{equation}\label{eq.nucH}
  \left(T^R + U_0^R \right) |\chi_{0 n}\> = E_{0 n} |\chi_{0 n}\>.
  \end{equation}
The analytical solution of this equation proceeds from the expansion of $U_0^R$ in terms of nuclear
displacements from their equilibrium coordinates $R_0$. By retaining terms up to quadratic order 
in the displacements and performing a linear transformation (\ref{app.phonons}) we find
the standard textbook result:
  \begin{equation}\label{eq.PES_expansion}
  U_0^R = U^{R_0}_0 + {\sum}_\nu \,\frac{1}{2} \, M_p \, \Omega_{\nu}^2 \, x_{ \nu}^2,
  \end{equation}
where $x_\nu$ and $\Omega_\nu$ are the amplitude and frequency of a vibrational mode $\nu$, 
and $M_p$ is a reference mass (in the following $x_\nu$ and $\Omega_\nu$ will only
be used in relation to the ground-state PES). The quadratic expansion in (\ref{eq.PES_expansion}) 
corresponds to the standard harmonic approximation for the ground state PES. This approximation has 
been very successful in many important cases~\cite{Baroni2001}, therefore it will be assumed
in the following. Accordingly we will consider temperatures well below the melting point 
of the solid or the dissociation energy of the molecule. The study of anharmonic corrections 
could be tackled using the approach described in~\cite{Monserrat20132}, however it will not 
be considered here since the formalism developed below is already rather involved.

After substituting (\ref{eq.PES_expansion}) into (\ref{eq.nucH}), the nuclear wavefunctions
$\chi_{0 n}(R)$  can be expressed as the product of one-dimensional quantum harmonic oscillators. 
In this case the state $|\chi_{0 n}\>$ is completely described by the set of integer occupation 
numbers of each quantized vibrational mode (\ref{app.phonons}). In the following the index $n$ 
in $|\chi_{0 n}\>$ will be understood to indicate the entire set of vibrational quantum numbers 
$n_\nu$. In the chemistry and in the solid-state literature the quanta of vibrational energy 
are referred to as ``vibrons'' and ``phonons'', respectively. In order to simplify the notation 
in the following we will use phonons to indicate such quanta, regardless of the system 
(molecule or solid).

\section{Heuristic approach to phonon-induced renormalization: temperature dependence and Allen-Heine theory}
\label{sec.heuristic}

In this section we analyse the effect of quantum nuclear dynamics on the electronic excitation
energies from a heuristic viewpoint. A more rigorous theory and its connection with the present 
derivation will be given in section \ref{sec.solids}.

If the nuclei could be held immobile in the configuration $R$, then the optical absorption
spectrum of the system introduced in section~\ref{sec.basicintro} would exhibit sharp peaks
at the energies $\e^R_\a$. The modification of these energies arising from the motion of the
nuclei around their equilibium configuration is referred to as ``phonon-induced renormalization''
and is discussed below.

If we assume that electronic transitions occur at fixed nuclear coordinates, then each configuration 
$R$ yields the excitation energies $\e^R_\a$. The probability of finding the nuclei in the configuration 
$R$ can be calculated in a first approximation using the nuclear wavefunctions in (\ref{eq.nucH}), 
and is given by $|\chi_{0 n}(R)|^2$. Therefore the transition energy averaged over all possible 
nuclear configurations can be obtained as $\int dR |\chi_{0 n}(R)|^2 \e^R_\a$, that is by evaluating 
the expectation value:
  \begin{equation}\label{eq.elec_exp}
  \<\e_\a\>_n = \<\chi_{0 n}|\e_\a^R|\chi_{0 n}\>.
  \end{equation}

At finite temperature $T$ the nuclear quantum states $|\chi_{0 n}\>$ will be occupied according 
to the Gibbs distribution law $\exp(-E_{0 n}/\kT)/Z$, where $Z = {\sum}_n \exp(-E_{0 n}/\kT)$ 
is the partition function and $k_\mathrm{B}$ Boltzmann's constant. Using these occupations we can 
evaluate the thermally-averaged excitation 
energy $\<\e_\a\>_T$ as:
  \begin{equation}\label{eq.semiclass}
  \<\e_\a\>_T = \frac{1}{Z} \sum_n e^{-\frac{E_{0 n}}{\kT} } \<\chi_{0 n}| \e_\a^R |\chi_{0 n}\>.
  \end{equation}
The heuristic argument used to derive this equation has been named the ``semiclassical Franck-Condon 
approximation'' in Ref.~\cite{Lax1952}.  The semiclassical approach has been used successfully in recent
first-principles calculations~\cite{Ramirez2006,Patrick2013}.
As we show in section~\ref{sec.mol}, the same result (\ref{eq.semiclass})
can be derived from the Franck-Condon theory.  

Equation (\ref{eq.semiclass}) is the starting point of the theory of temperature-dependent 
bandstructures developed by Allen and Heine~\cite{Allen1976}. In the Allen-Heine theory the evaluation 
of (\ref{eq.semiclass}) proceeds through the expansion of the electronic excitation energies in powers 
of the nuclear displacements from equilibrium (see \ref{app.phonons} for notation):
  \begin{equation}\label{eq.expansion}
  \e^R_{\a} = \e_\a + {\sum}_\nu \frac{\partial\e_\a}{\partial x_\nu}  x_\nu
  + \frac{1}{2} {\sum}_{\nu \mu} \frac{\partial^2\e_\a}{\partial x_\nu x_{\mu}} x_\nu x_{\mu} + \O(x_\nu^3).
  \end{equation}
By inserting (\ref{eq.expansion}) into (\ref{eq.semiclass}) and carrying out the integration
and the summation we obtain (\ref{app.phonons}):
  \begin{equation}\label{eq.AH_shift}
  \<\e_\a\>_T = \e_\a + {\sum}_\nu \, \frac{\partial \e_\a}{\partial n_\nu} \left[ n_\mathrm{B}(\Omega_\nu, T) 
              +  \frac{1}{2} \right] + \O(x_\nu^4),
  \end{equation}
where $n_{\rm B}(\Omega_\nu, T)= [\exp(\hbar\Omega_\nu/\kT) - 1]^{-1}$ is the Bose-Einstein distribution 
function (with $\hbar$ the Planck constant), and we introduced the electron-phonon coupling coefficient:
  \begin{equation}\label{eq.epcoupling}
  \frac{\partial \e_\a}{\partial n_\nu} = l_\nu^2
       \ \frac{\partial^2\e_\a}{\partial x_\nu^2}, 
  \end{equation}
with
  \begin{equation}\label{eq.epcoupling2}
  l_\nu = \sqrt{\frac{\hbar}{2M_p\Omega_\nu}}.
  \end{equation}
We note that this coupling coefficient contains both the ``Debye-Waller'' and the ``Fan'' terms of 
the Allen-Heine theory~\cite{Allen1976}.

For a single harmonic oscillator with frequency $\Omega_\nu$,
at high temperature ($\kT \gg \hbar\Omega_\nu$) (\ref{eq.AH_shift}) predicts
a linear dependence on temperature, while at low temperature
the characteristic zero-point effect becomes apparent. From (\ref{eq.AH_shift}) we see that the
``zero-point renormalization'' of the excitation energy is $(\partial \e_\a/\partial n_\nu)/2$.
In section~\ref{sec.Si} we use~(\ref{eq.AH_shift}) to calculate the temperature 
correction to the direct and indirect gaps of silicon.

Calculations based on the Allen-Heine theory have been employed successfully to describe the temperature
dependence of optical excitations in a number of materials~\cite{Giustino20102, 
  Marini2008, Gonze2011, Ramirez2006,Patrick2013, Allen1981, Allen1983, Capaz2005,Han2013,Monserrat2013,Cardona2005,Monserrat2014}.  
Nonetheless it is 
important to bear in mind that (\ref{eq.semiclass}), which underpins the Allen-Heine theory, constitutes 
a \emph{semiclassical} approximation. As a result, the Allen-Heine theory cannot resolve fine structures 
in optical spectra, and its accuracy is practically limited by the characteristic phonon energy of 
the system under consideration. This point will become more clear in sections~\ref{sec.mol_mom}~and
\ref{sec.ah_connection}, where we shall establish the connection between the semiclassical expression 
(\ref{eq.semiclass}) and a fully quantum-mechanical description of the electron-nuclear system.

\section{Perturbation theory: choice of the non-interacting Hamiltonian}\label{sec.pert}

Now we consider the complete Hamiltonian $\H$ of the joint electron-nuclear system, given in 
(\ref{eq.fullH}). We want to partition $\H$ into a non-interacting Hamiltonian $H_0$, for which exact 
formal solutions can be obtained, and perturbative corrections $\Delta H$, to be treated within
{\it time-independent} perturbation theory.

A perturbation approach is meaningful only when the corrections $\Delta H$ are small compared
to the non-interacting Hamiltonian $H_0$. This consideration leads naturally to two different
choices for $H_0$, one for molecules and one for extended solids. We start with the case of molecules.

\subsection{Non-interacting Hamiltonian and perturbation terms for molecules}\label{sec.hamiltonians}

In the case of molecules, defects in solids and Frenkel excitons, neutral excitations usually involve 
sizable variations of the electron density in a localized region of space, with a characteristic size 
of the order of a few bond lengths. In this case it is expected that the PES seen by the nuclei will be 
strongly dependent on the quantum state occupied by the electrons. This observation can be used to partition 
the complete Hamiltonian $\H$ as follows:
  \begin{eqnarray}\label{eq.part-mol}
  \H & =& H_0^\mathrm{m} + \Delta H^\mathrm{m},
  \end{eqnarray}
where ``m'' stands for ``molecular'', and the non-interacting Hamiltonian $H_0^\mathrm{m}$ is given by:
  \begin{equation}\label{eq.H0m}
  H_0^\mathrm{m} =  T^R +  {\sum}_\a U^R_\a  |\Psi_\a\>\<\Psi_\a|.
  \end{equation}
In this expression the PES $U^R_\a$ is defined through (\ref{eq.PESe}), and $\Psi_\a(r)$ denotes
a solution of~(\ref{eq.He}) for the equilibrium coordinates $R=R_0$. By construction 
the exact solutions of the non-interacting Hamiltonian $H_0^\mathrm{m}$ are given by $|\a n^{\rm m}\> = 
|\Psi_\a\> |\chi_{\a n}\>$, with $|\chi_{\a n}\>$ a solution of (\ref{eq.molnucH}). The exact 
eigenenergies of the non-interacting Hamiltonian are given by $E_{\a n}$ in (\ref{eq.molnucH}).
These properties can be verified directly by evaluating $H_0^\mathrm{m}|\a n^{\rm m}\>$. 
The factorized solution $|\a n^{\rm m}\>$ is nothing but a Born-Oppenheimer wavefunction~\cite{BornHuang}.

By combining (\ref{eq.fullH}), (\ref{eq.elecH}), (\ref{eq.PESe}), (\ref{eq.part-mol}) and (\ref{eq.H0m}), 
and using the completeness of the electronic states $|\Psi_\a\>$, we obtain the perturbative correction 
$\Delta H^\mathrm{m}$:
  \begin{eqnarray}
  \Delta H^\mathrm{m} = 
  \sum_\a \left[ \left(\<\Psi_\a|H_e^R|\Psi_\a\> - E_\a^R \right)|\Psi_\a\>\<\Psi_\a| 
  + \sum_{\b\neq\a} \<\Psi_\b|H_e^R|\Psi_\a\>|\Psi_\b\>\<\Psi_\a| \right]\!\!. \nonumber \\ \label{eq.molpert}
  \end{eqnarray}
In order to simplify the perturbation expansions in the following sections, it is useful
to express the first term in the square brackets in terms of the off-diagonal matrix elements 
$\<\Psi_\b|H_e^R|\Psi_\a\>$. This can be accomplished by expanding $E_\a^R$ in (\ref{eq.He})
in powers of $\left(H_e^R - H_e^{R_0}\right)$ (\ref{app.pert-exp}).
We find:
  \begin{equation}\label{eq.elec_en}
  \<\Psi_\a|H_e^R |\Psi_\a\> - E_\a^R = -
  \sum_{\b \neq \a}\frac{|\<\Psi_\b|H_e^R|\Psi_\a\> |^2}{E_\a - E_\b}+ \O(3).
  \end{equation}
Here and in the following the symbol $\O(n)$ means that the remaining terms are of the order 
of $\left(H_e^R-H_e^{R_0}\right)^n$. The rationale behind this expansion is discussed in~\ref{app.rcube}. 
If we define:
  \begin{equation}\label{eq.mat-el}
   V_{\b\a}^R = \<\Psi_\b|\left(H_e^R-H_e^{R_0}\right)|\Psi_\a\>,
  \end{equation}
we can rewrite (\ref{eq.molpert}) in the compact form:
  \begin{equation}\label{eq.ep-simple}
  \Delta H^\mathrm{m}  =
    -\sum_\a \sum_{\b \neq \a} \left[ \frac{|V_{\b\a}^R|^2}{E_\a - E_\b}
       |\Psi_\a\>\<\Psi_\a|  -  V_{\b\a}^R|\Psi_\b\>\<\Psi_\a| \right] + \O(3).
  \end{equation}
We will use this expression in evaluating the perturbative corrections in sections \ref{sec.molpert} 
and \ref{sec.molstates}.
 
\subsection{Non-interacting Hamiltonian and perturbation terms for solids}\label{sec.ham-solids}

In the case of an extended solid the choice of the non-interacting Hamiltonian made in the previous 
section is not optimal. For example, if we consider crystalline silicon and an excited state 
corresponding to a Wannier exciton extending over 5~nm, the charge density variation with respect 
to the ground state is less than $10^{-5}$~electrons/atom. Therefore, it is expected that the PES 
seen by the nuclei in this excited state will be essentially identical to that of the ground state.

Following this reasoning, it seems sensible to choose the non-interacting Hamiltonian so that 
its solutions only contain the nuclear wavefunctions $\chi_{0 n}$ from (\ref{eq.molnucH}),
corresponding to the ground-state PES $U_0^R$. This can be achieved by partitioning the complete 
Hamiltonian $\H$ as follows:
  \begin{eqnarray}\label{eq.part-sol}
  \H & =& H_0^\mathrm{s} + \Delta H^\mathrm{s},
  \end{eqnarray}
where ``s'' stands for ``solid-state'', and the non-interacting Hamiltonian $H_0^\mathrm{s}$ is:
  \begin{equation}\label{eq.H0s}
  H_0^\mathrm{s} =  T^R + U^R_0 + {\sum}_\a \e_\a |\Psi_\a\>\<\Psi_\a|.
  \end{equation}
The exact solutions of this Hamiltonian are $|\a n^\mathrm{s} \> = |\Psi_\a\> |\chi_{0 n}\>$, with energies 
$E_{\a n}^\mathrm{s} = \e_\a + E_{0 n}$. 
Since $\e_\a=E_\a^{R_0}-E_0^{R_0}$ vanishes for $\a = 0$, the solutions of $H_0^\mathrm{s}$ 
and $H_0^\mathrm{m}$ coincide in the ground state. 

Given the non-interacting Hamiltonian $H_0^\mathrm{s}$ in (\ref{eq.part-sol}), the associated perturbative 
correction reads:
  \begin{equation}\label{eq.solpert}
  \Delta H^\mathrm{s} = \Delta H^\mathrm{m} + \Delta H^\mathrm{AH}.
  \end{equation}
The term $\Delta H^\mathrm{m}$ is the same as in (\ref{eq.molpert}) and (\ref{eq.ep-simple}) 
for the molecular case, while the new term $\Delta H^\mathrm{AH}$ is given by:
  \begin{equation}\label{eq.AH}
  \Delta H^\mathrm{AH}   = {\sum}_\a (\e_\a^R - \e_\a)|\Psi_\a\>\<\Psi_\a|.  
  \end{equation}
The superscript ``AH'' stands for ``Allen-Heine'', and is meant to indicate that this term will
lead to the Allen-Heine theory of temperature-dependent band structures (section~\ref{sec.ah_connection}).
Again noting that $\e^R_0 =0$, the perturbation $\Delta H^\mathrm{AH}$ has no effect when the electrons are in their 
ground state. 

\section{Perturbative corrections to the non-interacting molecular Hamiltonian: Herzberg-Teller 
effect, Franck-Condon principle, and temperature dependence}\label{sec.mol}

We now derive the perturbative corrections to the energy and wavefunctions of the many-body 
electron-nuclear states $|\a n^{\rm m}\>$ and $|\a n^{\rm s}\>$ introduced in section~\ref{sec.pert}.
To this end we apply {\it time-independent} perturbation theory to the perturbations $\Delta H^\mathrm{m}$ 
and $\Delta H^\mathrm{s}$. \ref{app.pert-exp} provides a reminder of the general 
expressions for the perturbative expansions used below. In this section we start with the molecular 
Hamiltonian $H_0^\mathrm{m}$, while in sections~\ref{sec.solids}--\ref{sec.indir_solids} we consider 
the solid-state Hamiltonian $H_0^\mathrm{s}$.

\subsection{Perturbation corrections to the energies}\label{sec.molpert}

If we denote the exact energy of the joint electron-nuclear eigenstates of $\H$ by $E_{\a n}^{\rm e,m}$
(the superscript ``e'' standing for ``exact''), using (\ref{eq.en_exp}) and (\ref{eq.ep-simple}) we obtain:
  \begin{eqnarray}\label{eq.Emolfull}
  E_{\a n}^{\rm e,m} = E_{\a n} 
    - \sum_{\b\neq\a} \left[
      \frac{\<\chi_{\a n}|\, |V_{\a\b}^R|^2\,|\chi_{\a n}\>}{E_\a - E_\b}
    - \sum_m \frac{ |\<\chi_{\a n}| V_{\a\b}^R |\chi_{\b m}\>|^2}{E_{\a n} - E_{\b m} }
     \right] + \O(3). \nonumber \\
  \end{eqnarray}
This expression leads to a natural formal definition of the adiabatic approximation: in the following 
we will use the term {\it adiabatic approximation} in order to indicate the replacement:
  \begin{equation}\label{eq.adia}
  \sum_m \frac{ |\<\chi_{\a n}| V_{\a\b}^R |\chi_{\b m}\>|^2}{E_{\a n} - E_{\b m} }
  \simeq 
  \sum_m \frac{ |\<\chi_{\a n}| V_{\a\b}^R |\chi_{\b m}\>|^2}{E_{\a} - E_{\b} }.
  \end{equation}
This approximation is equivalent to stating that the electronic excitation energy $E_{\a} - E_{\b}$ is
much larger than the characteristic vibrational energy. To see this let us consider first the simplest 
scenario, whereby the potential energy surfaces $U_\a^R$ and $U_\b^R$ are shifted by a constant.
In this case, using the harmonic approximation, the operator identities given in \ref{app.phonons}, 
and a linear expansion of $V_{\a\b}^R$ in the atomic displacements, one finds that the matrix elements 
on the numerator only couple terms differing by one vibrational quantum number. As a result we have 
$E_{\a n} - E_{\b m} = E_{\a} - E_{\b} \pm \hbar\Omega$, with $\Omega$ the characteristic frequency 
associated with the potential energy surfaces. In more complicated situations, whereby the surfaces 
$U_\a^R$ and $U_\b^R$ differ by more than a constant, it is always possible to perform an expansion 
of $U_\b^R-U_\a^R$ in powers of $R$, and express $\chi_{\b m}$ in terms of the wavefunctions 
of $U_\a^R$ using perturbation theory. In this case the only terms appearing in the sum will 
be $m_\nu =n_\nu \pm 1$ (first-order expansion), $m_\nu =n_\nu \pm 2$ (second-order expansion), 
and so on, and the previous reasoning still applies. These observations show that (\ref{eq.adia}) 
will hold whenever $|E_{\a} - E_{\b}| \gg \hbar\Omega$, and hence corresponds to the usual statement 
of the adiabatic approximation (e.g.~\cite{Allen1976}). 

Using the adiabatic approximation defined by (\ref{eq.adia}) and the completeness relation 
$\sum_m |\chi_{\b m}\>\<\chi_{\b m}| = 1$, the term within the square brackets in (\ref{eq.Emolfull}) 
vanishes. This allows us to identify that term  with a non-adiabatic correction to the non-interacting
energy $E_{\a n}$, and rewrite (\ref{eq.Emolfull}) as:
  \begin{equation}\label{eq.mol-ad}
  E_{\a n}^{\rm e,m} = E_{\a n} + \O(3), \qquad \mbox{(a. a.)}
  \end{equation}
where ``(a. a.)'' stands for ``adiabatic approximation'' and reminds us that non-adiabatic terms 
are neglected. The most severe breakdown of (\ref{eq.adia}) will occur in the presence of degenerate 
electronic states. In this case the non-adiabatic coupling can be large and lead to a breakdown 
of the straightforward factorization of electronic and nuclear states~\cite{Domcke2012} (\ref{app.degen}). 

\subsection{Perturbation corrections to the states: the Born-Huang expansion}\label{sec.molstates}

We now use (\ref{eq.stat_exp}) to expand the exact electron-nuclear state $|\a n^{\rm e,m}\>$
in terms of the non-interacting Born-Oppenheimer states $|\a n\>$:
  \begin{equation}\label{eq.wfc_exp}
  |\a n^{\rm e,m}\> = |\Psi_\a\>|\chi_{\a n}\> + \sum_{\b\neq\a} 
   \sum_m  \frac{\<\chi_{\b m}| V_{\b\a}^R|\chi_{\a n}\>}{E_{\a n}  - E_{\b m} } 
   |\Psi_\b\> |\chi_{\b m}\> +  \O(2).
  \end{equation}
By working with the first order expansion we shall introduce errors of 
$\O(2)$  into the oscillator strengths.
On the other hand, by expanding the squared matrix elements it can be shown that these $\O(2)$ terms
only modify the strength of the zeroth order transition, and do not introduce any new features.
If we now apply the adiabatic approximation (\ref{eq.adia}) to the expansion in (\ref{eq.wfc_exp}), 
we obtain:
  \begin{equation}
  |\a n^{\rm e,m}\>\! =\! \left[ |\Psi_\a\> + \sum_{\b\neq\a}  \frac{ V_{\b\a}^R}{E_{\a}  - E_{\b} } |\Psi_\b\> 
   \right] \!|\chi_{\a n}\> 
  + \O(2).  \qquad \mbox{(a. a.)}
  \end{equation}
The expression within the brackets can be identified as the $R$-dependent electronic wavefunction
$\Psi_\a^R$. Expanding $|\Psi_\a^R\>$ in powers of $\left( H_e^R - H_e^{R_0}\right)$ 
we obtain (\ref{app.pert-exp}):
  \begin{equation}
  \label{eq.just_elec}
  |\Psi_\a^R\> = |\Psi_\a\> + \sum_{\b\neq\a} \frac{\<\Psi_\b|H_e^R | \Psi_\a\>}{E_\a - E_\b} 
     |\Psi_\b\> + \O(2).
  \end{equation}
Hence we obtain the following general expression for the joint electron-nuclear state:
  \begin{equation}\label{eq.BH}
  |\a n^{\rm e,m} \> = |\Psi_\a^R\>|\chi_{\a n}\> + \O(2). \qquad \mbox{(a. a.)} 
  \end{equation}
The quantity $|\Psi_\a^R\>|\chi_{\a n}\>$ appearing in (\ref{eq.BH}) is the leading term
in the so-called {\it Born-Huang expansion} of the joint electron-nuclear wavefunction~\cite{BornHuang}.
Therefore, our present analysis shows that the wavefunction $|\Psi_\a^R\>|\chi_{\a n}\>$ 
constitutes the first-order adiabatic approximation to the exact electron-nuclear wavefunction 
$|\a n^{\rm e,m} \> $ in molecules.

\subsection{Transitions: Herzberg-Teller effect and Franck-Condon principle}\label{sec.moltrans}

The perturbative expansions obtained in sections \ref{sec.molpert} and \ref{sec.molstates} will now 
be employed in order to analyze the expressions for the optical absorption spectra of molecules.
We consider an external driving perturbation $\Delta \cos(\w t)$, for example a uniform and oscillating 
electric field. This field induces transitions between the eigenstates of $\H$. The energy $\hbar\w$ 
is taken in the UV-Visible range, so that $\Delta$ can be considered to couple only to the electronic 
degrees of freedom, i.e.\ $\Delta = \Delta(r)$. We evaluate the transition rates 
$ W_{\a n \rightarrow \b m}^{\rm e,m}(\w)$ using the Fermi golden rule (considering absorption only): 
  \begin{equation}
  \label{eq.FGR}
  W_{\a n \rightarrow \b m}^{\rm e,m}(\w) = \frac{2\pi}{\hbar} |\< \b m^{\rm e,m}|\Delta|\a n^{\rm e,m}\>|^2
  \d(E^{\rm e,m}_{\b m} - E^{\rm e,m}_{\a n} -\hbar\omega).
  \end{equation}
Using (\ref{eq.mol-ad}) and (\ref{eq.BH}) we can rewrite the rates in terms of the Born-Oppenheimer
states $|\chi_{\a n}\>$ and $|\chi_{\b m}\>$:
  \begin{equation}\label{eq.HT_full}
  W_{\a n \rightarrow \b m}^{\rm e,m}(\w) = W_{\a n \rightarrow \b m}^\mathrm{HT}(\w)
  + \O(2,3), \qquad \mbox{(a. a.)}
  \end{equation}
having defined:
  \begin{equation}
  W_{\a n \rightarrow \b m}^\mathrm{HT}(\w) = \frac{2\pi}{\hbar} |\< \chi_{\b m} | P_{\b \a}^R |\chi_{\a n}\>|^2
  \d(E_{\b m} - E_{\a n} -\hbar\omega),
  \label{eq.HT}
  \end{equation}
and
  \begin{equation}\label{eq.P-def}
  P_{\b \a}^R = \<\Psi_\b^R|\Delta|\Psi_\a^R\>.
  \end{equation}
The notation $\O(n,m)$ in (\ref{eq.HT_full}) is used to indicate that the neglected terms are
of order $\O(n)$ in the strength of the transition and $\O(m)$ in the energy. 
We refer to the quantity $W_{\a n \rightarrow \b m}^\mathrm{HT}(\w)$ in (\ref{eq.HT}) as
the Herzberg-Teller rate, since it includes the characteristic $R$-dependence of the transition matrix 
element $P_{\b \a}^R$ known as the Herzberg-Teller effect~\cite{Herzbergbook}. 

The Herzberg-Teller rate can further be approximated by neglecting the dependence of the matrix 
element $P_{\b \a}^R$ on the nuclear coordinates. This corresponds to retaining only the zeroth 
order term in the expansion of the joint electron-nuclear state in (\ref{eq.wfc_exp}). We find:
  \begin{equation}
  W_{\a n \rightarrow \b m}^{\rm e,m}(\w) = W_{\a n \rightarrow \b m}^\mathrm{FC}(\w)
  + \O(1,3), \qquad \mbox{(a. a.)}
  \end{equation}
having defined:
  \begin{equation}
  W_{\a n \rightarrow \b m}^\mathrm{FC}(\w) = \frac{2\pi}{\hbar} |\< \chi_{\b m} | \chi_{\a n}\>|^2
  |P_{\b \a}|^2 \d(E_{\b m} - E_{\a n} -\hbar\omega),
  \label{eq.FC}
  \end{equation}
with $P_{\b \a} = P_{\b \a}^{R_0}$. This expression is the well-known rate of optical transitions 
according to the Franck-Condon principle~\cite{Atkinsbook}. In (\ref{eq.FC}) the electrons and nuclei
are completely decoupled, and the intensity of the transition results from the product of the 
electronic matrix element $P_{\b \a}$ and the overlap of the nuclear wavefunctions 
$\<\chi_{\b m}|\chi_{\a n}\>$.  The replacement of the Herzberg-Teller matrix element 
$P_{\b \a}^R$ by its zeroth-order approximation $P_{\b \a}$ is known as the Condon approximation.

The present analysis shows that the Herzberg-Teller rate and the Franck-Condon rate both correspond 
to the adiabatic approximation of the exact transition rate, correct to second order 
in the excitation energies. The strength of the transition is correct to first  order in the 
former, and to zeroth order in the latter.

\subsection{Temperature dependence: Connection to the Allen-Heine theory}\label{sec.mol_mom}

In this section we demonstrate the link between the Franck-Condon rate in (\ref{eq.FC}) and the 
thermally-averaged excitation energy $\<\e_\a\>_T$ introduced in (\ref{eq.semiclass}). To this end 
we evaluate the total rate of transitions from the electronic ground state $\Psi_0$ to the excited 
state $\Psi_\a$, irrespective of the vibrational quantum number of the final state. The temperature 
enters via the thermal distribution of the initial state among the vibrational quantum states 
$\chi_{0n}$, as in (\ref{eq.semiclass}):
  \begin{equation}
  W^\mathrm{FC}_{0 \rightarrow \a}(\w, T) = 
    \frac{1}{Z} \sum_n e^{-\frac{E_{0 n}}{\kT} } \, \sum_m W_{0 n \rightarrow \a m}^\mathrm{FC}(\w).
  \label{eq.fc-thermal-ave}
  \end{equation}
By using (\ref{eq.FC}) in this expression we find:
  \begin{equation*}
  W^\mathrm{FC}_{0 \rightarrow \a}(\w, T) =
  \frac{1}{Z} \sum_n e^{-\frac{E_{0 n}}{\kT} } \sum_m
  \frac{2\pi}{\hbar} |\< \chi_{\a m} | \chi_{0 n}\>|^2
  |P_{\a 0}|^2 \d(E_{\a m} - E_{0 n} -\hbar\omega).
  \end{equation*}
We can analyze this rate by inspecting the frequency moments. Using (\ref{eq.PESe}),
(\ref{eq.elec_excite}), (\ref{eq.molnucH}), as well as the completeness of the states $\chi_{\a m}$, the first moment can 
be written as:
  \begin{eqnarray} \label{eq.FC_mom}
  \<\hbar\omega\>^{{\rm FC},0\rightarrow \a}_T & = &  
   \int \!d\w \, \hbar\w \, W^\mathrm{FC}_{0 \rightarrow \a}(\w, T) \Big/
    \!\int \!d\w \, W^\mathrm{FC}_{0 \rightarrow \a}(\w, T) \nonumber \\
   & = & \frac{1}{Z} \sum_n e^{-\frac{E_{0 n}}{\kT} }
    \< \chi_{0 n} |\e_\a^R | \chi_{0 n}\> = \<\e_\a\>_T, 
  \end{eqnarray}
having used (\ref{eq.semiclass}) to obtain the last equality. This result indicates that the 
thermally-averaged excitation energy obtained heuristically in (\ref{eq.semiclass})--(\ref{eq.AH_shift}), 
which forms the basis for the Allen-Heine theory, corresponds to the first moment of the 
Franck-Condon lineshape. 

The comparison between the Franck-Condon lineshape and the Allen-Heine 
approach of section~\ref{sec.heuristic} 
can be extended to the case of higher frequency moments~\cite{Lax1952}. 
The width of the lineshape can be obtained from the second moment 
$\<\hbar^2\omega^2\>^{{\rm FC},0\rightarrow \a}_T$, and this quantity also matches
the square of $\e_\a^R$ in the Allen-Heine approach, i.e.   
\begin{eqnarray} \label{eq.FC_mom-2}
  \<\hbar^2\omega^2\>^{{\rm FC},0\rightarrow \a}_T =
  \<\e_\a^2\>_T = \frac{1}{Z} \sum_n e^{-\frac{E_{0 n}}{\kT} }
\<\chi_{0 n}|\left(\e_\a^R\right)^2 | \chi_{0 n}\>.
  \end{eqnarray} 
However, moving beyond the second moment introduces commutators
between the kinetic energy operator and potential energy surfaces~\cite{Lax1952}.
The analysis of frequency moments will be investigated in further detail in section~\ref{sec.SC}.

In summary, the present analysis indicates that, on the one hand, the Allen-Heine approach outlined
in section~\ref{sec.heuristic} can be rooted on a solid ground by starting from perturbation
theory and invoking the Franck-Condon approximation. On the other hand, it also indicates that
the Allen-Heine expression (\ref{eq.AH_shift}) represents only an {\it average} excitation
energy.
As a consequence, the Allen-Heine theory is inadequate for resolving 
fine structures in optical spectra or small energy differences, and caution should be used when comparing 
to experiment.

This section concludes our discussion of electron-vibration coupling in molecules; 
in sections~\ref{sec.solids}--\ref{sec.indir_solids} we will 
build on the results obtained so far to
discuss electron-phonon coupling 
in solids.

\section{Perturbative corrections to the non-interacting solid-state Hamiltonian: electron-phonon 
renormalization and Allen-Heine theory}\label{sec.solids}

In the previous section we applied perturbation theory to the non-interacting molecular Hamiltonian
$H_0^\mathrm{m}$, and established the connection to the Herzberg-Teller and Franck-Condon expressions 
for optical absorption in molecules, (\ref{eq.HT}) and (\ref{eq.FC}).
Now we repeat the perturbation theory analysis for the solid-state picture, starting from the 
non-interacting solid-state Hamiltonian $H_0^\mathrm{s}$ (\ref{eq.H0s}). In this section
we establish the link between the perturbative corrections to the energy levels and
the electron-phonon renormalization resulting from the Allen-Heine theory~\cite{Allen1976}.
In section~\ref{sec.indir_solids} we will investigate the perturbative corrections to the joint
electron-nuclear wavefunctions, and analyse phonon-assisted absorption in solids.

\subsection{Perturbation corrections to the energies}\label{sec.sol_energy}

Similarly to section~\ref{sec.molpert} we denote the exact energy of the joint electron-nuclear 
states of $\H$ by $E_{\a n}^{\rm e,s}$. Using (\ref{eq.ep-simple})--(\ref{eq.AH}) with (\ref{eq.en_exp})
we find:
  \begin{eqnarray}\label{eq.esol_exp}
  E_{\a n}^{\rm e,s} &=& E_{\a n}^{\rm s} + \Delta E_{\a n}^{\rm AH} + \O(3) \nonumber \\
  && -\sum_{\b \neq \a}
  \left[ \frac{\<\chi_{0 n}| |V_{\b\a}^R|^2|\chi_{0 n}\>}{E_\a - E_\b} -
  \sum_m \frac{ |\<\chi_{0 n}| V_{\a\b}^R |\chi_{0 m}\>|^2 }{E_{\a} - E_{\b} + E_{0 n} - E_{0 m} }\right],   
  \end{eqnarray}
where $\Delta E_{\a n}^\mathrm{AH}$ is given by:
  \begin{eqnarray}\label{eq.energy-ah}
  \Delta E_{\a n}^{\rm AH} =  \<\chi_{0 n}| \e_\a^R|\chi_{0 n}\> -\e_\a
  + \!\sum_{m\neq n}\frac{\left| \<\chi_{0 m}|\e_\a^R + 
   \left[ \<\Psi_\a | H_e^R  |\Psi_\a\>-\!E_\a^R
   \right] |\chi_{0 n}\>\right|^2}{E_{0 n} - E_{0 m}}. \nonumber \\
  \end{eqnarray}
The term appearing within square brackets in (\ref{eq.esol_exp}) is the solid-state counterpart 
of the non-adiabatic corrections already discussed for molecules [see (\ref{eq.Emolfull})].
This term can be neglected when the vibrational contribution $E_{0 n} - E_{0 m}$ in the second term
is small compared to the electronic excitation energy $E_{\a} - E_{\b}$. This observation
allows us to state a formal definition of the adiabatic approximation in solids as follows:
  \begin{equation}\label{eq.adia_s}
  \sum_m \frac{ |\<\chi_{0 n}| V_{\a\b}^R |\chi_{0 m}\>|^2 }{E_{\a} - E_{\b} + E_{0 n} - E_{0 m} } \simeq
   \sum_m \frac{ |\<\chi_{0 n}| V_{\a\b}^R |\chi_{0 m}\>|^2 }{E_{\a} - E_{\b}}.  
  \end{equation}
This approximation constitutes the analogue of (\ref{eq.adia}) for the case of solids. 
Also in this case it is expected that (\ref{eq.adia_s}) will hold for systems with an electronic 
energy gap on the electronvolt scale. Under this conditions the square brackets in (\ref{eq.esol_exp})
can safely be ignored.

\subsection{The zero-point energy}

The energy correction $\Delta E_{\a n}^{\rm AH}$ appearing in (\ref{eq.energy-ah})
can be written in a more intuitive form by recasting all the $R$-dependent quantities
in terms of displacements along the vibrational eigenmodes of the system. By proceeding
along the same lines as in (\ref{eq.expansion})--(\ref{eq.epcoupling}) and using the
algebra of ladder operators in \ref{app.phonons} we obtain:
  \begin{eqnarray}
  \Delta E_{\a n}^{\rm AH} = 
  &=& \sum_\nu \, \frac{\partial \e_\a}{\partial n_\nu} \left( n_\nu +  \frac{1}{2} \right)
  - \sum_\nu \frac{l_\nu^2}{\hbar\Omega_\nu} \left(\frac{\partial\e_\a}{\partial x_\nu}\right)^{\!\!2}
   + \O(x_\nu^4). 
  \end{eqnarray}
In order to reach this expression it is convenient to express the second sum in (\ref{eq.energy-ah})
using (\ref{eq.elec_en}) and (\ref{eq.mat-el}), and observe that the cross terms in the square
modulus vanish since they couple $|\chi_{0 n_\nu}\>$ with $|\chi_{0,n_\nu\pm 1}\>$ and $|\chi_{0,n_\nu\pm 2}\>$,
respectively. 
If we define the zero-point renormalization $E^\mathrm{ZP}_{\a}$ as the part of the energy-correction 
independent of the phonon numbers $n_\nu$:
  \begin{equation}
  \label{eq.ZP}
  E^\mathrm{ZP}_{\a} = \frac{1}{2} \sum_\nu \, \frac{\partial \e_\a}{\partial n_\nu} 
  - \sum_\nu \frac{l_\nu^2}{\hbar\Omega_\nu} \left(\frac{\partial\e_\a}{\partial x_\nu}\right)^{\!\!2},
  \end{equation}
we can rewrite (\ref{eq.esol_exp}) more compactly as:
  \begin{equation}
  E_{\a n}^{\rm e,s} = E_{\a n}^{\rm s} + E_\a^\mathrm{ZP} 
        + \sum_\nu \frac{\partial \e_\a}{\partial n_\nu} n_\nu +  \O(3). \qquad \mbox{(a. a.)}
  \label{eq.solid_corr}
  \end{equation}
This expression, which was derived using second-order perturbation theory starting from
the many-body electron-nuclear wavefunctions, is similar but not identical to the Allen-Heine
result (\ref{eq.AH_shift}). In fact our expression contains an additional contribution
to the zero-point energy, that is the second term in (\ref{eq.ZP}). Such contribution arises
from the second sum in (\ref{eq.energy-ah}), and is absent in the Allen-Heine theory since
the latter neglects off-diagonal couplings between different vibrational wavefunctions.

So far first-principles as well as semiempirical calculations based on the Allen-Heine 
theory~\cite{Giustino20102,Cannuccia2011,Gonze2011,Cardona2005} did not include the extra term in (\ref{eq.ZP}).
However, it will be important for future calculations to establish the magnitude of this correction.

\section{Perturbative corrections to the solid-state non-interacting Hamiltonian:
zero-phonon absorption and phonon-assisted absorption}\label{sec.indir_solids}

\subsection{Perturbation corrections to the states and optical matrix elements}\label{sec.solstates}

In order to obtain the optical transition rates we use Fermi's golden rule as in
section~\ref{sec.moltrans}. The counterpart of (\ref{eq.FGR}) for solids is trivially: 
  \begin{eqnarray} 
  W^{\rm e,s}_{\a n \rightarrow \b m} (\omega) &=& \frac{2\pi}{\hbar}
  | \<\b m^{\rm e,s}|\Delta|\a n^{\rm e,s}\>|^2 
  \, \delta(E^{\rm e,s}_{\b m} - E^{\rm e,s}_{\a n} - \hbar \omega).
  \label{eq.wholerate-exact}
  \end{eqnarray}
The perturbative corrections to the energies appearing in the delta function were obtained 
in section~\ref{sec.sol_energy}. Here we need to derive the perturbative expansion of the joint 
electron-nuclear wavefunctions in order to evalate the optical matrix elements 
$\<\b m^{\rm e,s}|\Delta|\a n^{\rm e,s}\>$. Using (\ref{eq.ep-simple}), (\ref{eq.solpert}) and (\ref{eq.AH})
 inside (\ref{eq.stat_exp}) we find:
  \begin{eqnarray}
   |\a n^{\rm e,s}\> &= &|\a n^{\rm s}\> 
  + \sum_{m\neq n}\frac{\<\chi_{0 m}|\e_\a^R |\chi_{0 n}\>}{E_{0 n} - E_{0 m} } |\a m^{\rm s}\>  \nonumber \\
  && + \sum_{\b\neq\a} \sum_m \frac{\<\chi_{0 m}|V_{\b\a}^R|\chi_{0 n}\>}{E_\a - E_\b + E_{0 n} - E_{0 m}} 
   |\b m^{\rm s}\> + \O(2).
  \label{eq.sol_noadia_exp}
  \end{eqnarray}
After replacing the expansion (\ref{eq.sol_noadia_exp}) inside the transition matrix elements 
we obtain:
  \begin{eqnarray}
    \<\b m^{\rm e,s}|\Delta|\a n^{\rm e,s}\> = \Delta_{\a n,\b m}^{\rm dir,NP} + 
       \Delta_{\a n,\b m}^{\rm dir,PA} 
       + \Delta_{\a n,\b m}^{\rm ind,NP} 
       + \Delta_{\a n,\b m}^{\rm ind,PA} +\O(2),
  \label{eq.wholerate}
  \end{eqnarray}
with the definitions:
  \begin{eqnarray}
   \Delta_{\a n,\b m}^{\rm dir,NP} =  P_{\b \a} \,\d_{nm}, \label{eq.dir} \\
   \Delta_{\a n,\b m}^{\rm dir,PA}  =  P_{\b \a} (1-\d_{nm}) \frac{\<\chi_{0 m}|(\e_\a^R - \e_\b^R )
      |\chi_{0 n}\>}{E_{0 n} - E_{0 m} }, \label{eq.dir-PA}\\
  \Delta_{\a n,\b m}^{\rm ind,NP}  = \d_{nm}\left[
  \sum_{\gamma\ne \b} 
  \frac{\<\chi_{0 n}|V_{\b\gamma}^R|\chi_{0 n}\> P_{\gamma\a}}{\e_\b - \e_\gamma}
  +\sum_{\gamma\ne \a} 
  \frac{P_{\b\gamma}\<\chi_{0 n}|V_{\gamma\a}^R|\chi_{0 n}\>}{\e_\a - \e_\gamma}
  \right],
  \label{eq.ind} \\
  \Delta_{\a n,\b m}^{\rm ind,PA}  =  (1-\d_{nm}) \left[
  \sum_{\gamma\ne \b} 
  \frac{\<\chi_{0 m}|V_{\b\gamma}^R|\chi_{0 n}\> P_{\gamma\a}}{\e_\b \!-\! \e_\gamma + E_{0 m} \!-\! E_{0 n}}
  +\sum_{\gamma\ne \a} 
  \frac{P_{\b\gamma}\<\chi_{0 m}|V_{\gamma\a}^R|\chi_{0 n}\>}{\e_\a \!-\! \e_\gamma + E_{0 n} \!-\! E_{0 m}}
   \right]\!. \nonumber \\ \label{eq.ind-PA} 
  \end{eqnarray}
The partitioning of the optical matrix element in (\ref{eq.wholerate}) naturally leads to the
identification of {\it direct} and {\it indirect} transitions. In direct transitions
the optical matrix element between electron-only wavefunctions is allowed, i.e.\ $P_{\b\a}\ne 0$.
In this case we find ``no-phonon'' transitions [with the superscript ``dir,NP'', (\ref{eq.dir})],
whereby the nuclear quantum number does not change, and ``phonon-assisted'' transitions [with 
the superscript ``dir,PA'', in (\ref{eq.dir-PA})], where the initial and final nuclear states differ.
In indirect transitions the optical matrix element $P_{\b\a}$ vanishes, but there can be
a contribution to the oscillator strength coming from the interactions between electrons
and vibrations. This is indicated as ``ind'' in (\ref{eq.ind-PA}). Also in this case we can further 
distinguish between no-phonon (\ref{eq.ind}) and phonon-assisted (\ref{eq.ind-PA}) indirect transitions. 
However,  this no-phonon contribution is non-vanishing only for quadratic order in displacements, while
the phonon-assisted contributions contribute at linear order. Accordingly we will not discuss (\ref{eq.ind}) further. 
As will be clear in the following sections, the transitions associated with $\Delta_{\a n,\b m}^{\rm dir,NP}$
correspond to the standard optical absorption in direct gap semiconductors (e.g.\ GaAs), while those
associated with $\Delta_{\a n,\b m}^{\rm ind,PA}$ correspond to the onset of indirect gap
semiconductors (e.g.\ Si).

At this stage we can expand $\e_\a^R$ and $V_{\a\b}^R$ in terms 
of nuclear displacements around the equilibrium configuration, as in (\ref{eq.expansion}). To this end 
we define the ``many-body electron-phonon matrix element'':
  \begin{equation}\label{eq.bigG}
  G^\nu_{\a \b} = l_\nu \frac{\partial V_{\a\b}^R}{\partial x_\nu},
   \end{equation}
with $l_\nu$ and $V_{\b\gamma}$ given by (\ref{eq.epcoupling2}) and (\ref{eq.mat-el}), respectively.
Using this definition and the transformation to normal mode coordinates (\ref{app.phonons}) we have:
  \begin{eqnarray}
  V_{\a\b}^R &=& \sum_\nu G^\nu_{\a \b}(b_\nu^\dagger + b_\nu) + \O(x_\nu^2),\label{eq.V-x} \\
  \e_\a^R &=& \e_\a + \sum_\nu G^\nu_{\a \a}(b_\nu^\dagger + b_\nu) + \O(x_\nu^2), \label{eq.epsa-x}
  \end{eqnarray}
where $b^\dagger_\nu$ and $b_\nu$ are the standard raising and lowering operators, respectively,
see (\ref{eq.x-b}). The last expression was obtained from (\ref{eq.He}), (\ref{eq.mat-el}) 
and (\ref{eq.bigG}). This change of coordinates 
allows us to simplify (\ref{eq.dir})--(\ref{eq.ind-PA}) and identify important selection rules. 
For example, using (\ref{eq.epsa-x}) and (\ref{eq.V-x}) the matrix elements appearing in (\ref{eq.dir-PA}) 
and (\ref{eq.ind-PA}) become:
   \begin{eqnarray}
   (1-\d_{nm})\<\chi_{0 m}|\e_\a^R |\chi_{0 n}\> &= &
      \sum_\nu G^\nu_{\a \a} 
    \left( \sqrt{n_\nu+1} \,\d_{m_\nu,n_\nu+1} + \sqrt{n_\nu} \, \d_{m_\nu,n_\nu-1} \right)
      \nonumber \\ &&\hspace{5cm} + \O(x_\nu^2), \label{eq.dir-PA-matel}\\
   (1-\d_{nm}) \<\chi_{0 m}|V_{\a\b}^R|\chi_{0 n}\> &=& 
    \sum_\nu G^\nu_{\a \b}  
    \left( \sqrt{n_\nu+1} \,\d_{m_\nu,n_\nu+1} + \sqrt{n_\nu} \, \d_{m_\nu,n_\nu-1} \right)
      \nonumber \\ &&\hspace{5cm}+ \O(x_\nu^2), \label{eq.ind-PA-matel}
   \end{eqnarray}
where the sets of integers $n_\nu$ and $m_\nu$ identify the occupations of each normal mode
in the quantum states $|\chi_{0 n}\>$ and $|\chi_{0 m}\>$, respectively. The Kronecker delta
$\d_{m_\nu,n_\nu\pm 1}$ is meant to be 1 when all the modes in $|\chi_{0 n}\>$ and $|\chi_{0 m}\>$
have the same number of phonons, except mode $\nu$, for which $m_\nu = n_\nu\pm 1$.
Clearly the matrix elements in (\ref{eq.dir-PA-matel})--(\ref{eq.ind-PA-matel}) are associated 
with the standard concepts of phonon absorption ($m_\nu = n_\nu+1$) and emission ($m_\nu = n_\nu-1$).

Our final observation regarding the optical matrix elements is to note that the transition rates in 
(\ref{eq.wholerate-exact}) require the square modulus of (\ref{eq.wholerate}),
which can be written as:
  \begin{eqnarray}
  |\<\b m^{\rm e,s}|\Delta|\a n^{\rm e,s}\>|^2 = |\Delta_{\a n,\b m}^{\rm dir,NP}|^2 +
    |\Delta_{\a n,\b m}^{\rm dir,PA}|^2 
     + |\Delta_{\a n,\b m}^{\rm ind,PA}|^2 +\O(2).
     \label{eq.dip2}
  \end{eqnarray}
 By inspecting 
the factors $\d_{nm}$ and $(1-\d_{nm})$ in (\ref{eq.dir})--(\ref{eq.ind-PA}) and considering 
that in the adiabatic approximation
$|\Delta_{\a n,\b m}^{\rm ind,NP}|\ll |\Delta_{\a n,\b m}^{\rm dir,NP}|$ and 
$|\Delta_{\a n,\b m}^{\rm ind,PA}|\ll |\Delta_{\a n,\b m}^{\rm dir,PA}|$, we find
we find that the $\O(2)$ cross terms provide a much smaller contribution.
Therefore we evaluate the transition rates corresponding to the
various processes identified in (\ref{eq.dir}), (\ref{eq.dir-PA}), and (\ref{eq.ind-PA})
separately,
starting with the no-phonon direct transition rate.

\subsection{No-phonon direct transitions: Connection to Allen-Heine theory}
\label{sec.ah_connection}

Using  (\ref{eq.wholerate-exact}) and (\ref{eq.dip2})  it is natural to define the no-phonon direct 
transition rate as:
  \begin{equation}
  W^{\rm dir,NP}_{\a n \rightarrow \b m} (\omega) = \frac{2\pi}{\hbar}
  | \Delta_{\a n,\b m}^{\rm dir,NP} |^2
  \, \delta(E^{\rm e,s}_{\b m} - E^{\rm e,s}_{\a n} - \hbar \omega).
  \end{equation}
After inserting  (\ref{eq.solid_corr}) and (\ref{eq.dir})   into this expression we obtain:
  \begin{eqnarray}
  W^{\rm dir,NP}_{\a n \rightarrow \b m} (\omega) = \d_{nm} \frac{2\pi}{\hbar}| P_{\b \a} |^2 \times\nonumber \\
  \hspace{0.7cm} \delta \left[
   (\e_\b-\e_\a) + (E_\b^\mathrm{ZP}-E_\a^\mathrm{ZP}) 
    + \sum_\nu \left(\frac{\partial \e_\b}{\partial n_\nu} - \frac{\partial \e_\a}{\partial n_\nu} \right) 
   n_\nu - \hbar \omega \right]  + \O(3), \nonumber \\ \hspace{11cm} \mbox{(a. a.)}
  \end{eqnarray}
where $\O(3)$ refers to the energy expansion. This rate describes the optical transitions found 
at the absorption onset of direct gap semiconductors. The key difference with standard expressions 
found in the literature~\cite{Bassani_book} is that here we have additional structure arising from
the nuclear motion. This is best seen by considering the thermal average of the rate of transitions 
from the fundamental state, as in (\ref{eq.fc-thermal-ave}):
  \begin{eqnarray}
   W^{\rm dir,NP}_{0 \rightarrow \a}(\w, T) = \frac{2\pi}{\hbar}| P_{\a 0} |^2 
    \frac{1}{Z} \sum_n e^{-\frac{E_{0 n}}{\kT} } 
 \delta\left[ \e_\a + E_\a^\mathrm{ZP} + \sum_\nu \frac{\partial \e_\a}{\partial n_\nu} 
   n_\nu - \hbar \omega \right] \nonumber \\ 
   \hspace{8.5cm}+ \O(3).  \qquad \mbox{(a. a.)} \label{eq.dir-NP-lineshape}
\label{eq.AH_trans_temp}
  \end{eqnarray}
A thorough discussion of the lineshape $W^{\rm dir,NP}_{0 \rightarrow \a}(\w, T)$ will be presented 
in section~\ref{sec.urbach}. For now we simply point out that the absorption onset, as described 
by the first moment of the lineshape, is easily calculated from (\ref{eq.dir-NP-lineshape}) as it was 
already done in (\ref{eq.FC_mom}). We find:
  \begin{eqnarray}
  \<\hbar\omega^{\rm dir,NP}_{0 \rightarrow \a}\>_T = \<\e_\a\>_T 
  - \sum_\nu \frac{l_\nu^2}{\hbar \Omega_\nu} \left(\frac{\partial\e_\a}{\partial x_\nu}\right)^2,
  \label{eq.dir-NP-onset}
  \end{eqnarray}
where $\<\e_\a\>_T$ is the same as in (\ref{eq.AH_shift}).  By comparing (\ref{eq.dir-NP-onset}) 
and (\ref{eq.FC_mom}) we realize that the thermal average of the absorption onset calculated using 
the Allen-Heine theory, i.e.\ $\<\e_\a\>_T$ from (\ref{eq.AH_shift}), although coinciding with the 
average of the Franck-Condon offset [$\<\hbar\omega^{\rm FC}_{0 \rightarrow \a}\>_T$ in
(\ref{eq.FC_mom})], differs slightly from the direct no-phonon onset, owing to the presence of 
an extra term in (\ref{eq.dir-NP-onset}).  This extra term does not appear in the original theory~\cite{Allen1976}.
In view of performing accurate comparisons between theory and experiment it will be important to 
establish the magnitude of this additional term in first-principles calculations.

\subsection{Phonon-assisted direct transitions}\label{sec.phon-assist-direct}

In the same spirit of section~\ref{sec.ah_connection} we use the partitioning in (\ref{eq.dip2})
to define the ``phonon-assisted direct transition rate'':
  \begin{equation}
  W^{\rm dir,PA}_{\a n \rightarrow \b m} (\omega) = \frac{2\pi}{\hbar}
  | \Delta_{\a n,\b m}^{\rm dir,PA} |^2 
  \, \delta(E^{\rm e,s}_{\b m} - E^{\rm e,s}_{\a n} - \hbar \omega).
  \end{equation}
Using (\ref{eq.dir-PA}), (\ref{eq.solid_corr}), and (\ref{eq.dir-PA-matel}) in this expression we obtain:
  \begin{eqnarray}
  W^{\rm dir,PA}_{\a n \rightarrow \b m} (\omega) = \frac{2\pi}{\hbar}
  \sum_{\nu,\pm} 
   \left|\frac{P_{\b\a}(G^\nu_{\a \a}-G^\nu_{\b \b})}{\hbar\Omega_\nu}\right|^2 
  \d_{m_\nu,n_\nu \pm 1}  
  \left[n_\nu+\frac{1}{2}\pm\frac{1}{2}\right] 
  \nonumber \\
   \hspace{4cm} \times \delta \left( \e_\b - \e_\a \pm \hbar\Omega_\nu + \Delta\e_{\b\a,n}^{\nu\pm} -\hbar\w
  \right) + \O(x_\nu^4,3), \nonumber \\ \hspace{10.5cm} \mbox{(a. a.)} \label{eq.direct}
  \end{eqnarray}
where we have defined the energy correction $\Delta\e_{\b\a,n}^{\nu\pm}$ as follows:
  \begin{equation}
  \Delta\e_{\b\a,n}^{\nu\pm} = (E_\b^\mathrm{ZP}-E_\a^\mathrm{ZP})
        + {\sum}_\mu 
      \left( \frac{\partial \e_\b}{\partial n_\mu} -\frac{\partial \e_\a}{\partial n_\mu} \right)\!n_\mu\,
        \pm \frac{\partial \e_\b}{\partial n_\nu}.
       \label{eq.en-corr}
  \end{equation}
We can gain some intuition on the meaning 
of (\ref{eq.direct}) by considering the simplest possible scenario, whereby the initial state 
$\Psi_\alpha$ is the electronic ground state, there is only one phonon in the mode of frequency 
$\Omega_\nu$, i.e.\ $n_\nu=1$, and the energy correction in (\ref{eq.en-corr}) is negligible.
In this case (\ref{eq.direct}) yields two transitions. One transition corresponds to the creation 
of an additional phonon in the system, so that in the final state $n_\nu=2$. In this case the excitation 
energy is larger than the direct band gap $\e_\b -\e_\a$. The second transition corresponds to the destruction
of a phonon, with the final states having $n_\nu=0$. In this case the transition energy is smaller 
than the direct band gap by the amount $\hbar\Omega_\nu$.  This indicates that it is possible for 
the system to make transitions below the optical gap by sourcing the extra energy from the phonon 
bath. This phenomenon corresponds to {\it phonon-assisted sub-gap absorption}.
Since sub-gap absorption is allowed only when the electronic system can source the missing 
energy from the phonon bath, this phenomenon is only possible when the system is above 
its zero-point state, i.e.\  sub-gap absorption cannot occur at $T=0$. 

More generally it is possible to study the temperature dependence of phonon-assisted
direct absorption onsets along the lines of (\ref{eq.AH_trans_temp}) and (\ref{eq.dir-NP-onset}).
The additional complication with respect to no-phonon transitions is that the intensity
of phonon-assisted transitions in (\ref{eq.direct}) is modulated by the phonon numbers.
Concentrating on transitions from the ground state we find:
  \begin{eqnarray}
   W^{\rm dir,PA}_{0\rightarrow \a} (\omega,T) = \frac{2\pi}{\hbar}
   \frac{1}{Z} \sum_n e^{-\frac{E_{0 n}}{\kT} }
   \sum_{\nu,\pm}   \left| \frac{P_{\a 0}G^\nu_{\a \a}}{\hbar\Omega_\nu} \right|^2
   \left[n_\nu+\frac{1}{2}\pm\frac{1}{2}\right] \nonumber \\
   \hspace{3.6cm} \times \,\delta\!\left( \e_\a \pm \hbar\Omega_\nu + \Delta\e_{\a 0,n}^{\nu\pm} -\hbar\w \right)
    + \O(x_\nu^4,3). \,\,\,\mbox{(a. a.)} \label{eq.temp-dir-PA}
   \end{eqnarray}
 Further aspects 
of temperature-dependent lineshapes will be investigated in section~\ref{sec.urbach}.

Sub-gap phonon-assisted transitions had already been proposed many decades ago using 
a time-dependent perturbation theory description of indirect absorption~\cite{Dumke1957}. Yet, to the best
of our knowledge, this proposal was not followed up, and such contributions to the optical spectra 
have not been included in first-principles calculations. The possibility of sub-gap absorption 
should be taken in serious consideration when trying to compare the calculated band gap renormalization
of solids with optical experiments~\cite{Giustino20102,Marini2008,Gonze2011}, since they can offset the 
calculated onset by as much as a phonon energy.

\subsection{Phonon-assisted indirect transitions}\label{sec.indir_sol_bigG}

In this section we conclude the analysis started in sections~\ref{sec.ah_connection} and 
\ref{sec.phon-assist-direct} by considering the oscillator strength associated with the matrix 
elements in (\ref{eq.ind-PA}). We define the ``phonon-assisted indirect transition rate'':
  \begin{equation}
  W^{\rm ind,PA}_{\a n \rightarrow \b m} (\omega) = \frac{2\pi}{\hbar}
  | \Delta_{\a n,\b m}^{\rm ind,PA} |^2
  \, \delta(E^{\rm e,s}_{\b m} - E^{\rm e,s}_{\a n} - \hbar \omega).
  \end{equation}
Using (\ref{eq.solid_corr}), (\ref{eq.ind-PA}),  (\ref{eq.ind-PA-matel}), and (\ref{eq.en-corr}) 
this can be rewritten as:
  \begin{eqnarray}
  W^{\rm ind,PA}_{\a n \rightarrow \b m} (\omega) = \frac{2\pi}{\hbar}
     \sum_{\nu,\pm}  \d_{m_\nu,n_\nu\pm 1} 
     \left| 
    \sum_{{\gamma}\ne \b} \frac{ G^\nu_{\b {\gamma}}P_{{\gamma}\a}
    }{\e_\b \!-\! \e_{\gamma} \pm \hbar\Omega_\nu}
  + \sum_{{\gamma}\ne \a} \frac{ P_{\b {\gamma}}G_{{\gamma}\a}^\nu
    }{\e_\a \!-\! \e_{\gamma} \mp \hbar\Omega_\nu}
     \right|^2 \nonumber \\
   \hspace{2.22cm} \times \left[n_\nu+\frac{1}{2}\pm\frac{1}{2}\right] 
     \delta \left( \e_\b - \e_\a \pm \hbar\Omega_\nu + \Delta\e_{\b\a,n}^{\nu\pm} -\hbar\w \right) \nonumber\\
     \hspace{6cm} + \O(x_\nu^4,3). \qquad \mbox{(a. a.)} 
  \label{eq.ind-all}
  \end{eqnarray}
We can stress the similarity with standard textbook expressions by replacing $\e_\b$ in the
first denominator using the argument of the Dirac delta function. This replacement yields:
  \begin{eqnarray}
  W^{\rm ind,PA}_{\a n \rightarrow \b m} (\omega) = \frac{2\pi}{\hbar}
     \sum_{\nu,\pm}  \d_{m_\nu,n_\nu\pm 1} \left[n_\nu+\frac{1}{2}\pm\frac{1}{2}\right] \nonumber \\
    \hspace{2.2cm} \times \left| \sum_{{\gamma}\ne \b} \frac{ G^\nu_{\b {\gamma}}P_{{\gamma}\a}
    }{\e_\gamma - \e_\a + \Delta\e_{\b\a,n}^{\nu\pm}-\hbar\w}
  + \sum_{{\gamma}\ne \a} \frac{ P_{\b {\gamma}}G_{{\gamma}\a}^\nu
    }{\e_\gamma -\e_\a \pm \hbar\Omega_\nu}
     \right|^2 \nonumber \\
   \hspace{2.2cm} \times 
     \delta \left( \e_\b - \e_\a \pm \hbar\Omega_\nu + \Delta\e_{\b\a,n}^{\nu\pm} -\hbar\w \right) 
     + \O({x_\nu^4},3). \,\,\,\mbox{(a. a.)} 
  \label{eq.ind-all2}
  \end{eqnarray}
This expression is almost identical to those derived in Ref.~\cite{Bassani_book} and used in 
Ref.~\cite{Noffsinger2012} for calculating the indirect absorption edge of Si from first principles.
The only difference between our present formulation and that of Ref.~\cite{Bassani_book}
is that here the electron-phonon renormalization is included through the energy correction
$\Delta\e_{\b\a,n}^{\nu\pm}$. Neglecting such renormalization leads exactly to the usual 
expression for indirect absorption~\cite{Bassani_book}.

The two amplitudes appearing in (\ref{eq.ind-all2}) are traditionally interpreted as corresponding 
to the successive absorption of a photon and of a phonon ($G^\nu_{\b \gamma} P_{\gamma \a}$),
and vice versa ($P_{\b \gamma} G^\nu_{\gamma \a}$). The forms used in the corresponding denominators
are meant to mimic the energy selection rules associated with these two processes. However
it should be stressed that these are so-called ``virtual'' transitions, therefore the shapes
of the denominators are more mnemonic expedients rather than actual selection rules.

One interesting point to be highlighted is that, while most investigations of indirect absorption
in solids describe the phonon bath as a {\it time-dependent} perturbation to the electronic 
system, in our case electrons and vibrations are treated on the same footing. In our approach
the only time-dependent potential is that of the external field, while phonons are described
using {\it time-independent} perturbation theory, consistently with the notion that vibrations
exist at all times in the system. In this way we do not need to assume an artificial
adiabatic switch-on of the electron-phonon interaction in the distant past~\cite{Sakuraibook}.

The present formulation carries important implications for practical first-principles calculations
of indirect absorption including electron-phonon renormalization. In fact, (\ref{eq.ind-all})
shows clearly that the effects of temperature and zero-point energy shifts should be included
{\it only} in the transition energies [through (\ref{eq.en-corr})], and {\it not} in the denominators 
of the transition amplitudes. This aspect is important in order to ensure a consistent
description of electron-phonon interactions and phonon-assisted absorption. Without the present theory 
there would be significant ambiguity as to where and how to include such energy shifts in the formalism.

Also in this case it is possible to study the temperature-dependent lineshape at the absorption onset 
by performing a thermal average, precisely as in (\ref{eq.temp-dir-PA}). Here we refrain from giving 
the complete expression since it is almost identical to (\ref{eq.temp-dir-PA}), the only change being the
replacement of the square modulus with the one appearing in (\ref{eq.ind-all}).

\section{Exponential lineshapes and the Urbach tail}\label{sec.urbach}

The temperature-dependent absorption lineshapes in (\ref{eq.AH_trans_temp}) and (\ref{eq.temp-dir-PA})
exhibit a shift of the absorption peaks which is proportional to the phonon quantum numbers $n_\nu$
through the energy corrections $\Delta\e_{\b\a,n}^{\nu\pm}$ in (\ref{eq.en-corr}). In addition, in the case 
of phonon-assisted absorption, also the intensity of each transition is modulated by the phonon quantum 
numbers via the factors $n_\nu+1$ or $n_\nu$ [see (\ref{eq.temp-dir-PA}); the same holds for indirect 
absorption in section~\ref{sec.indir_sol_bigG}]. In this section we show how these effects lead to 
an absorption lineshape which is surprisingly similar to the famous Urbach tail~\cite{Kurik1971}.

\begin{figure*}
  \centering
  \includegraphics[height=60mm]{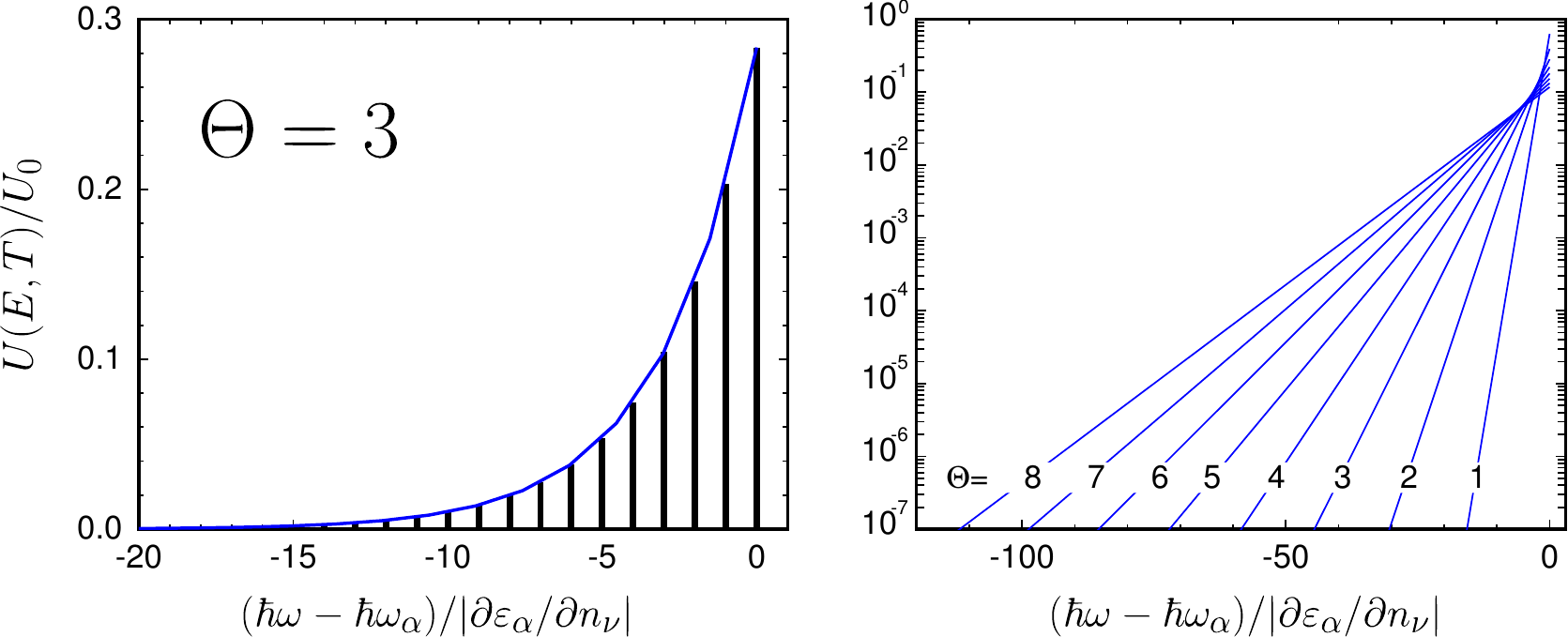}
  \caption{\label{fig.urbach}
  Envelope $U(E,T)$ of the absorption 
  lineshapes $W^\mathrm{dir,NP}_{0 \rightarrow \a}(\w,T)$ as derived in (\protect\ref{eq.my_urbach}). 
  This function is meant to represents the exponential tail in the optical absorption spectrum 
  just below the direct gap $\hbar\w_\a$ of a solid. (a) Exponential decay of the absorption rate 
  at the temperature $\Theta = 3$, with $\Theta = \kT/\hbar\Omega_\nu$ (blue curve). The vertical 
  black lines represent the intensities of the Dirac delta functions in (\protect\ref{eq.urb1}).
  (b) The absorption lineshapes plotted on a logarithmic scale at various temperatures, $\Theta = 1,\dots,8$. 
  We see that the lineshapes yield straight lines and radiate approximately from a common focus, 
  as prescribed by the Urbach rule~\protect\cite{Kurik1971}.}
  \end{figure*}

In order to illustrate this effect we consider the simplest situation, corresponding to no-phonon 
direct transitions as in (\ref{eq.AH_trans_temp}) and a single vibrational mode; a qualitatively 
similar behavior is found in all the other cases. From (\ref{eq.AH_trans_temp}) we have:
  \begin{eqnarray}
   W^{\rm dir,NP}_{0 \rightarrow \a}(\w, T) = A_\nu(T)
     \sum_{n_\nu} \exp\left[-\frac{\hbar\Omega_\nu}{\kT}n_\nu\right]
 \,\delta\!\left[ \hbar\w_\a + \frac{\partial \e_\a}{\partial n_\nu} n_\nu - \hbar \omega \right]
  \label{eq.urb1}
  \end{eqnarray}
with $\hbar\w_{\a} = \e_\a + E^\mathrm{ZP}_\a$ and $A_\nu(T) = (2\pi/\hbar)| P_{\a 0} |^2
/[n_B(\Omega_\nu,T)+1]$. If we consider the direct gap of tetrahedral semiconductors as an example, 
the coefficient $\partial \e_\a/\partial n_\nu$ corresponding to the highest optical phonons at the centre
of the Brillouin zone will be negative~\cite{Cardona2005}. In this case the lineshape in 
(\ref{eq.urb1}) will exhibit peaks {\it below} the direct gap, equally spaced by the energy 
$\partial \e_\a/\partial n_\nu$, with a strength decreasing exponentially as one moves away 
from the edge.  This exponential lineshape is shown in figure~\ref{fig.urbach}, and is tentatively 
identified as the {\it Urbach tail}~\cite{Urbach1953, Kurik1971}. 

By connecting the maxima of the absorption peaks in (\ref{eq.urb1}) we obtain immediately
the envelope of the lineshape:
  \begin{eqnarray}
  U(\w,T) = \frac{U_0}{n_B(\Omega_\nu,T) + 1} \ \exp \left[ \frac{\hbar\w - \hbar\w_\a}{E_\nu(T)} \right],
  \label{eq.my_urbach}
  \end{eqnarray}
with $U_0$ a temperature-independent constant, and the decay parameter $E_\nu$ given by:
  \begin{equation} \label{eq.urb-decay}
  E_\nu(T) = \left|\frac{\partial\e_\a}{\partial n_\nu}\right| \frac{\kT}{\hbar\Omega_\nu}.
  \end{equation}
Our expression for the exponential tail bears a very strong resemblance to that derived 
{\it empirically} over a wide range of materials~\cite{Kurik1971}:
  \begin{equation}
  U^\mathrm{emp}(E,T) =U^\mathrm{emp}_0 \exp \left[ \frac{\hbar\w - \hbar\omega_\a}
     {E^{\rm emp}(T)} \right],
  \label{eq.full_urbach}
  \end{equation}
with 
  \begin{equation}\label{eq.urb-decay-exp}
  E^{\rm emp}(T) = E_0^{\rm emp} \left[n_B(\Omega^\mathrm{emp}_\nu,T) + \frac{1}{2} \right],
  \end{equation}
and $U^\mathrm{emp}_0$, $\Omega^\mathrm{emp}_\nu$, and $E_0^{\rm emp}$ experimentally-determined constants. In particular, 
the similarity between our result (\ref{eq.my_urbach}) and the empirical observation (\ref{eq.full_urbach}) 
at high temperature is striking. In fact, for $\kT \gg \hbar\Omega_\nu$ our temperature prefactor 
and the empirical one do coincide, since $n_B(\Omega_\nu,T) + 1/2 \simeq \kT/\hbar\Omega_\nu$.
Furthermore, as shown in figure~\ref{fig.urbach} our theory predicts that the lineshapes obtained 
at any temperature appear to radiate from a common focus. This behavior is one of the characteristic 
traits of the Urbach rule~\cite{Kurik1971}.

The electron-phonon coupling coefficient appearing in (\ref{eq.urb-decay}) is typically 
of the order of the phonon energy~\cite{Cardona2005}, therefore our theory predicts a decay parameter 
of the order of $\kT$. This finding is consistent with experimental measurements of band tails 
in a variety of solids~\cite{Kurik1971}.

The key qualitative difference between our present theory and empirical observations of exponential 
absorption edge is that our width vanishes at $T=0$, while (\ref{eq.urb-decay-exp}) remains finite. 
One possibility to explain such a discrepancy is to assume that additional temperature-independent 
mechanisms may cause some broadening which has been incorporated empirically in (\ref{eq.urb-decay-exp}) 
and is not taken into account in our formalism. However, before making any claims it will be 
important to carry out detailed first-principles calculations, and compare {\it quantitative} 
numerical predictions of exponential tails with the available experimental data.

Attempts to link the Urbach tail to electron-phonon interactions date back to some 
of the earliest theoretical work on the subject~\cite{Toyozawa1959,Keil1966,Davydov1968}.
However, to the best of our knowledge this is the first time that the Urbach tail has been
derived entirely from first principles, and found to be connected to the theory of phonon-induced 
renormalization in solids.

\section{Unifying expressions for molecules and solids in a semiclassical approach}
\label{sec.SC}
\subsection{Motivation}

In the previous sections we discussed two different viewpoints of the electron-phonon problem,
the molecular picture and the solid-state picture.
The reader will note that the expressions derived starting from the molecular picture
are in general less complicated than their solid-state counterparts.
For instance, the expression for the Herzberg-Teller rate (\ref{eq.HT}) is
far more compact than the solid-state partitioning of (\ref{eq.dip2})
into
no-phonon, phonon-assisted, direct and indirect contributions.
This behaviour can be attributed to the fact that solutions to the non-interacting Hamiltonian
in the molecular picture (\ref{eq.H0m}) carry information about the excited state PES,
while in the solid-state case this information must be recovered \emph{a posteriori} through the perturbative
term $\Delta H^\mathrm{AH}$ (\ref{eq.AH}).

It is sensible therefore to ask whether the molecular picture can be applied
to extended solid-state systems.
The difficulty here is that calculating the $\a$-dependent nuclear
wavefunctions through (\ref{eq.molnucH})  requires knowledge of
excited-state forces, which are highly nontrivial to calculate 
in extended systems~\cite{Beigi2003}.
An ideal compromise would be to find an approach which inherits the basic structure
of the molecular picture but nonetheless avoids explicit evaluation of the $\a$-dependent
nuclear wavefunctions.
In this section we discuss such an approach.

\subsection{Semiclassical approximation for electron-phonon 
renormalization and optical absorption}

Here we consider the semiclassical approach originally proposed by 
Lax in Ref.~\cite{Lax1952} as an alternative
to the Herzberg-Teller expression (\ref{eq.HT}) in section~\ref{sec.moltrans}. 
Its derivation proceeds by expressing the absorption rate appearing in (\ref{eq.HT}) in the time domain,
  \begin{equation}
  W_{\a n \rightarrow \b m}^\mathrm{HT}(\w) = 
  \frac{1}{\hbar} \int dt e^{-i\w t} W_{\a n \rightarrow \b m}^\mathrm{H T}(t),
  \label{eq.FT}
  \end{equation}
which gives:
  \begin{eqnarray}
  W_{\a n \rightarrow \b m}^\mathrm{HT}(t) = \frac{2\pi}{\hbar} 
   \<\chi_{\a n}|P_{\a\b}^R|\chi_{\b m}\> \nonumber \\
    \hspace{2.7cm} \times \<\chi_{\b m}|
     \exp\left(\frac{it}{\hbar}E_{\b m}\right) 
    P_{\b\a}^R
     \exp\left(-\frac{it}{\hbar}E_{\a n}\right) 
   |\chi_{\a n}\>.
  \end{eqnarray}
As the nuclear wavefunctions are eigenstates of (\ref{eq.molnucH}), we can replace
the eigenvalues by the operators to find:
  \begin{eqnarray}
  W_{\a n \rightarrow \b m}^\mathrm{HT}(t)& =& \frac{2\pi}{\hbar} 
  \<\chi_{\a n}|P_{\a\b}^R|\chi_{\b m}\> \nonumber \\
  && \times\! \<\chi_{\b m}|\exp\left[\frac{it}{\hbar} (T^R\! + U_\b^R )\right]
  \!P_{\b\a}^R \!\exp\left[-\frac{it}{\hbar} (T^R\! + U_\a^R )\right] |\chi_{\a n}\>. \nonumber \\
  \label{eq.timedomain}
  \end{eqnarray}
The semiclassical approximation by Lax consists of neglecting all commutators involving
the kinetic energy operator (e.g.\ $[T^R,U_\a^R]$, $[T^R,P_{\b\a}^R]$ and so on). Using this 
simplification we can rewrite
(\ref{eq.timedomain}) as follows:
  \begin{eqnarray}
  W_{\a n \rightarrow \b m}^\mathrm{HT}(t)& \simeq & \frac{2\pi}{\hbar} 
  \<\chi_{\a n}|P_{\a\b}^R|\chi_{\b m}\> 
   \<\chi_{\b m}|\exp\left[\frac{it}{\hbar} (\e_\b^R-\e_\a^R) \right]
  \!P_{\b\a}^R |\chi_{\a n}\>
  \label{eq.semiclasstimedomain}
  \end{eqnarray}
where we used (\ref{eq.PESe}) and (\ref{eq.elec_excite}) to 
rewrite $U_\b^R - U_\a^R = \e_\b^R-\e_\a^R$.
Now using (\ref{eq.FT}) to return to the frequency domain,
summing over all possible final vibrational states, and setting
the initial electronic state to the ground state, we obtain the simple expression:
  \begin{eqnarray}\label{eq.W-sc}
  W^{\rm SC}_{0 n \rightarrow \a} (\omega) =  \<\chi_{0 n}| W^R_{0\rightarrow \a} (\omega) |\chi_{0 n}\>,
  \end{eqnarray}
with
  \begin{eqnarray}\label{eq.W-sc-2}
  W^R_{0\rightarrow \a} (\omega) = \frac{2\pi}{\hbar} |P_{\a 0}^R|^2 \delta( \e_\a^R - \hbar \omega).
  \end{eqnarray}
The expression (\ref{eq.W-sc}) gives the optical absorption spectrum in terms of the average 
over the ground-state nuclear wavefunctions $\chi_{0 n}(R)$ of the absorption spectra  
$W^R_{0\rightarrow \a}$
obtained for nuclei immobile in the configurations $R$.
In (\ref{eq.W-sc-2}) the electron-phonon interaction is taken into account via 
$\e_\a^R$, $P_{\a 0}^R$, and $\chi_{0 n}(R)$, while the use of the relation
$\sum_m |\chi_{\a m}\> \<\chi_{\a m}| = 1$ has removed the problematic 
$\a$-dependent nuclear wavefunctions.
The approximation defined by (\ref{eq.W-sc}) is referred to as ``semiclassical'' since 
it becomes exact in the limit where the nuclei are so heavy that the spectrum of the harmonic oscillator 
becomes continuous.

Performing the thermal average of (\ref{eq.W-sc}) gives the temperature dependent semiclassical expression,
  \begin{equation}
  W^\mathrm{SC}_{0 \rightarrow \a}(\w, T) =
    \frac{1}{Z} {\sum}_n e^{-\frac{E_{0 n}}{\kT} } \, W_{0 n \rightarrow \a}^\mathrm{SC}(\w).
  \label{eq.sc-thermal-ave}
  \end{equation}
This expression can be simplified further if we use Mehler's formula~\cite{Watson1933}. In fact, after 
combining (\ref{eq.W-sc-2}), (\ref{eq.sc-thermal-ave}), and (\ref{eq.mehler}) 
we obtain the compact result:
  \begin{equation}\label{eq.sc-mehler-T}
  W^\mathrm{SC}(\w,T) = \int {\prod}_\nu dx_\nu 
     \frac{1}{\sqrt{2\pi\<x_\nu^2\>_T}} 
     \exp\left(-\frac{x_\nu^2}{2\<x_\nu^2\>_T}\right) W^R_{0\rightarrow \a} (\omega),
  \end{equation}
with $R$ and $x_\nu$ related as in (\ref{eq.u-R}). This result has a simple intuitive interpretation: 
in the semiclassical approximation the temperature-dependent optical absorption spectrum is obtained 
by first calculating spectra for nuclei clamped in a variety of configurations $R$, and then averaging 
the spectra thus obtained using a gaussian importance function. The width of the importance function 
increases with the temperature as $\<x_\nu^2\>_T = l_\nu^2 \, [2 n_{\rm B}(\Omega_\nu,T) + 1]$ 
(\ref{app.phonons}).

An appealing aspect of the method proposed in this section is that it can be used without
difficulty with any electronic structure package which can compute optical absorption spectra
at fixed nuclei, without requiring a significant investment in software development.
For example, in~\cite{Patrick2013} we computed (\ref{eq.W-sc}) for diamondoids using Importance 
Sampling Monte Carlo integration. Alternatively, Path-Integral Monte Carlo techniques can also 
be employed~\cite{DellaSala2004,Schwartz2009}.

\subsection{Connection to the Herzberg-Teller effect in molecules and indirect absorption in solids}

We stress that the approximation leading to (\ref{eq.semiclasstimedomain}) is purely {\it heuristic},
and the validity of the ensuing formulation should be assessed by comparing with the predictions
of the complete theory. Here we limit ourselves to the analysis of the first frequency moment
of the lineshape; a more comprehensive discussion can be found in Ref.~\cite{Lax1952}.
By proceeding along the lines of section~\ref{sec.mol_mom} we find the following expression
relating the first frequency moment of the ``exact'' Herzberg-Teller lineshape, (\ref{eq.HT}), 
and the ``approximate'' semiclassical lineshape, (\ref{eq.W-sc}):
  \begin{eqnarray}\label{eq.SC-HT-moment}
   \< \hbar\w \>^{\rm HT}_{0 n \rightarrow \a} =
         \< \hbar\w \>^{\rm SC}_{0 n \rightarrow \a} + \frac{\< \chi_{0 n}| P_{0\a}^R[T^R, P_{\a 0}^R] 
          |\chi_{0 n}\>}{\< \chi_{0 n}| |P_{\a 0}^R|^2 |\chi_{0 n}\>}.
  \end{eqnarray}
Analogous relations are found for higher frequency moments. The magnitude of the last term in
(\ref{eq.SC-HT-moment}) can be estimated from the linear expansion of $P_{0\a}^R$ as
  \begin{equation*}
   \< \hbar\w \>^{\rm HT}_{0 n \rightarrow \a} - \< \hbar\w \>^{\rm SC}_{0 n \rightarrow \a} \sim
    \sum_\nu \hbar\Omega_\nu \left|
  \sum_{\gamma\neq \a} \frac{G^\nu_{\a \gamma} P_{\gamma 0} }{\e_\a - \e_\gamma}
  +\sum_{\gamma\neq 0} \frac{ P_{\a \gamma} G^\nu_{\gamma 0}}{\e_0 - \e_\gamma}
  \right|^2 \frac{1}{|P_{0\a}|^2}.
  \end{equation*}
This result indicates that the semiclassical lineshape is expected to capture very accurately 
the first moment of the complete Herzberg-Teller lineshape, since the error is a fraction
$(G/E_{\rm g})^2$ of the characteristic vibrational energy, with $G$ a typical electron-phonon
matrix element and $E_{\rm g}$ the fundamental gap.

In the case of solids it is possible to perform a similar analysis and show that
the semiclassical approximation in (\ref{eq.W-sc}) correctly captures the onset
of direct and indirect absorption, both in terms of transition energies and oscillator
strengths. As an example we consider here the oscillator strength for indirect
absorption, which is obtained from (\ref{eq.ind-all}) as:
      \begin{eqnarray}\label{eq.osc-st-ind}
    \int d\w\, W^{\rm ind,PA}_{0 n \rightarrow \a} (\omega) = \frac{2\pi}{\hbar}
     \sum_\nu
     \left|
    \sum_{{\gamma}\ne \a} \frac{ G^\nu_{\a {\gamma}}P_{{\gamma} 0}
    }{\e_\a \!-\! \e_{\gamma} }
  + \sum_{{\gamma}\ne 0} \frac{ P_{\a {\gamma}}G_{{\gamma} 0}^\nu
    }{\e_0 \!-\! \e_{\gamma} }
     \right|^2 \!\!(2n_\nu+1)\nonumber \\
     \hspace{6cm} + \O(x_\nu^4,3). \qquad \mbox{(a. a.)}
  \end{eqnarray}
In order to reach this expression we used the fact
that in the adiabatic approximation the vibrational energies 
are small with respect to the fundamental gap, $\hbar\Omega_\nu \ll E_{\rm g}$.
The semiclassical counterpart of (\ref{eq.osc-st-ind}) is obtained from (\ref{eq.W-sc})
and (\ref{eq.W-sc-2}):
  \begin{eqnarray}\label{eq.sc-indir}
  \int d\w \, W^{\rm SC}_{0 n \rightarrow \a} (\omega) = 
  \frac{2\pi}{\hbar} \<\chi_{0 n}| |P_{\a 0}^R|^2 |\chi_{0 n}\>.
  \end{eqnarray}
By expanding the optical matrix element $P_{\a 0}^R$ in this expression about the equilibrium positions 
of the nuclei using (\ref{eq.just_elec}) and (\ref{eq.V-x}), and setting
$P_{0 \a}=0$ (indirect process) we find:
  \begin{equation}\label{eq.dP-ladder}
  P_{0 \a}^R = 
  \sum_\nu\left[
  \sum_{\gamma\neq 0} \frac{G^\nu_{0 \gamma} P_{\gamma \a} }{\e_0 - \e_\gamma}
  +\sum_{\gamma\neq\a} \frac{ P_{0 \gamma} G^\nu_{\gamma \a}}{\e_\a - \e_\gamma}
  \right] (b_\nu^\dagger + b_\nu) + \O(x_\nu^2).
  \end{equation}
The replacement of this expansion inside (\ref{eq.sc-indir}) yields, after using the
standard algebra of ladder operators (\ref{app.phonons}):
  \begin{eqnarray}
  \int d\w \, W^{\rm SC}_{0 n \rightarrow \a} (\omega) =
    \int d\w\, W^{\rm ind,PA}_{0 n \rightarrow \a} (\omega)  + \O(x_\nu^4).
  \end{eqnarray}
This result indicates that the semiclassical approximation 
correctly captures the oscillator strength of indirect optical transitions in solids.
A similar reasoning applies to direct transitions.

\subsection{Advantages and shortcomings of the semiclassical approximation}

The semiclassical approach defined by (\ref{eq.sc-thermal-ave}) carries the advantage
of starting from the more accurate molecular non-interacting Hamiltonian 
without needing information about the $\a$-dependent
nuclear wavefunction.
As such, it provides a unified framework 
for studying solids and molecules using exactly the same formalism and the same computational 
techniques. This aspect is especially important given the large volume of research activity in
the areas of nanoscience and nanotechnology, where one is often confronted with heterogeneous
systems, e.g.\ molecular adsorbates on surfaces.

The main shortcoming of the semiclassical approach in molecules is that the characteristic
Franck-Condon structure consisting of distinct vibronic peaks is completely lost.
In fact, as the numerical tests of Ref.~\cite{Keil1965} demonstrate, the neglect of the commutators 
in (\ref{eq.timedomain}) destroys precisely the quantisation of the vibrational energy levels. 
In practice the semiclassical approximation in molecules is very useful for calculating the 
{\it envelope} of the absorption profile, without resolving individual vibronic transitions. 
Additionally the semiclassical approximation is expected to improve as the size of the 
molecule increases.  This is clearly demonstrated in our earlier work on 
diamondoids~\cite{Patrick2013}, where we showed 
that in the case of triamantane (C$_{18}$H$_{24}$) the semiclassical approach yields excellent 
agreement with experiment.

Apart from the practical advantage of (\ref{eq.sc-thermal-ave}) only 
relying on the nuclear wavefunctions $\chi_{0 n}$ in the electronic ground 
state, an additional strength is found by noting that the approach 
requires  neither the {\it harmonic} approximation nor the 
{\it adiabatic} approximation to be satisfied by the excited states.
This observation is supported by empirical evidence: the model 
calculations of Ref.~\cite{Bersukerbook} demonstrate that the semiclassical expression can capture 
non-adiabatic Jahn-Teller effects; in addition, our calculations of the optical spectra 
of adamantane within the semiclassical approach~\cite{Patrick2013} are in excellent agreement 
with experiment, even though this molecule has a triply degenerate highest-occupied molecular 
orbital and undergoes a Jahn-Teller splitting upon excitation~\cite{Patzer2012} (\ref{app.degen}).

In summary the semiclassical approach seems to offer a useful compromise between
computational simplicity, accuracy, and broad applicability to the widest range of systems.
First-principles calculations will be needed to carry out a systematic assessment of the
performance of this method in reproducing experimental spectra.
In the following section we demonstrate the application
of (\ref{eq.sc-thermal-ave}) to the calculation of the optical absorption
spectrum of bulk silicon.

\section{The semiclassical approximation applied to bulk silicon}\label{sec.Si}

\subsection{Introduction}
In this section we apply some of the expressions derived above to
the prototypical indirect gap semiconductor, bulk silicon.
As noted in the introduction to this manuscript, there has been
phenomenal progress in the developments of electronic structure
methods for dealing with the many-electron problem~\cite{Onida2002}.
Here we shall work at the 
level of the local density approximation to density-functional theory.
Although such calculations generally fail to obtain quantitative
agreement with experimental observations (most famously underestimating
the band gap), qualitative features can be reproduced.
As discussed in section~\ref{sec.indir_sol_littleg},
the calculation of electron-phonon renormalization and phonon-assisted
optical absorption using more complicated electronic structure methods
is an important subject for future research.

\subsection{Computational approach}

Here we describe the technical details of our calculations.
The reader interested in results may choose to skip to section~\ref{sec.si_results}.

\subsubsection{Electronic structure}

We calculate the energy-level renormalization and
optical absorption spectra using~(\ref{eq.semiclass})
and (\ref{eq.sc-thermal-ave}).
We replace
the electronic excitation energies $\e_\a$ appearing in these
equations with Kohn-Sham eigenvalues obtained within
the local-density approximation to DFT~\cite{Hohenberg1964,Kohn1965}.
Similarly the many-body matrix elements $P_{\b\a}^R$ (\ref{eq.P-def}) are 
replaced by those taken between single-particle Kohn-Sham wavefunctions.

\subsubsection{Energy-level renormalization}

We evaluate the energy-level renormalization in two ways.
In the first case, we obtain the electron-phonon coupling coefficients
$\partial\e_\a/\partial n_\nu$ by averaging the energies $\e_\a^R$
calculated after displacing the nuclei by amplitudes $\pm \sqrt(\hbar/M_p\Omega_\nu)$
along a phonon mode $\nu$, which isolates the quadratic term in the expansion
(\ref{eq.expansion})~\cite{Capaz2005}.
Substituting the calculated coefficients into (\ref{eq.AH_shift})
yields the temperature-dependent energy $\<\e_\a\>_T$.
Alternatively, we can obtain $\<\e_\a\>_T$ directly from the analogue
of (\ref{eq.sc-mehler-T}), i.e.:
  \begin{equation}\label{eq.sc-mehler-energy}
  \<\e_\a\>_T = \int {\prod}_\nu dx_\nu 
     \frac{1}{\sqrt{2\pi\<x_\nu^2\>_T}} 
     \exp\left(-\frac{x_\nu^2}{2\<x_\nu^2\>_T}\right) \e_\a^R.
  \end{equation}
Calculating $\<\e_\a\>_T$ in this way allows us to assess
the impact of neglecting the $x_\nu^4$ terms in (\ref{eq.AH_shift}).
In the case that eigenvalues are degenerate at the equilibrium
structure, we evaluate $\<\e_\a\>_T$ as a trace (\ref{app.degen}).

\subsubsection{Absorption spectra}
In order to compare our absorption spectrum to experiment, we construct
the absorption coefficient $\kappa$ as
\begin{equation}
\kappa(\omega,T) = \frac{\omega}{c} \frac{\epsilon_2(\w,T)}{n^r},
\end{equation}
where $n^r$ is the refractive index.
Here for simplicity we neglect the frequency and temperature dependence of the
refractive index, and use the experimental value $n^r=3.4$~\cite{Green1995}.
The temperature-dependent imaginary part of the dielectric
function $\epsilon_2(\w,T)$ is found by noting that, for clamped ions
$\epsilon^R_2(\w) \propto \ \sum_\a 1/\w \ W^R_{0\rightarrow \a}(\w)$;
therefore in the semiclassical picture (\ref{eq.sc-thermal-ave})
we obtain
\begin{equation}
\label{eq.sc-mehler-eps}
\epsilon_2(\w,T) = \int {\prod}_\nu dx_\nu 
     \frac{1}{\sqrt{2\pi\<x_\nu^2\>_T}} 
     \exp\left(-\frac{x_\nu^2}{2\<x_\nu^2\>_T}\right) \epsilon_2^R(\w)
\end{equation}
In practice we use the momentum representation of the matrix elements
to evaluate  $\epsilon^R_2(\w)$, and neglect the commutator term
arising from the nonlocal part of the pseudopotential~\cite{Baroni1986}.
We replace the $\delta$-functions which appear in $\epsilon^R_2(\w)$ 
with Gaussians of width 0.2~eV for the spectrum obtained with fixed ions and 0.02~eV
for the semiclassical calculation.
Finally, in order to account to the band gap problem, we impose a rigid scissor
shift of 0.7~eV to the energies of the unoccupied Kohn-Sham states,
obtained as the difference between DFT and $GW$ calculations in Ref.~\cite{Lambert2013}.

\subsubsection{Sampling method}

In both (\ref{eq.sc-mehler-energy}) and (\ref{eq.sc-mehler-eps})
we must evaluate an integral over all nuclear displacements.
In a previous work~\cite{Patrick2013} we recast the integral as a sum
over a large sample of nuclear geometries, generated such that
the phonon displacements $x_\nu$  were distributed according to
the Gaussian factor  $\exp\left(-x_\nu^2/2\<x_\nu^2\>_T\right)$.
Here we repeat that general approach, but employ the method described
in Ref.~\cite{Brown2013} to generate the sample.
This method replaces uniformly-distributed random numbers in the generation
algorithm with a low-discrepancy (Sobol) sequence, which greatly improves
the convergence properties in higher-dimensional systems~\cite{MC}.
We used sample sizes of 200 steps to compute both the energy-level renormalization
and optical spectrum.
We tested the convergence of the former by increasing the sample size to 500 steps,
and found the calculated corrections to change by less than 2~meV.

\subsubsection{Computational details}

Electronic structure calculations were performed within
the local-density approximation to DFT, using
plane-wave basis sets and periodic boundary conditions 
implemented in the \texttt{Quantum ESPRESSO} 
distribution~\cite{quantumespresso}.
We used a norm-conserving pseudopotential~\cite{Fuchs1999} to descibe the Si
ion and expanded the electronic wavefunctions in reciprocal
space up to an energy cutoff of 35~Ry.
For the calculation of the energy-level renormalization we
used a $4\times4\times4$ supercell and sampled the electrons
at the $\Gamma$-point, while for the optical
spectrum we used a  $2\times2\times2$ supercell with an $8\times8\times8$
Brillouin Zone sampling of the electrons (i.e.\ an effective electronic 
sampling of $16\times16\times16$).
The equilibrium structures were calculated by varying the lattice
parameter $a$ until the force on each atom was less than 0.03~eV/\AA \ 
and the 
pressure less than 0.5~kbar, yielding values of 5.41 and 5.31~\AA \ for
the $4\times4\times4$/$2\times2\times2$ supercells.
The phonon modes were determined by displacing each ion
in the primitive cell by 0.005~\AA, obtaining the forces, then using the translational
symmetry and appropriate sum rules~\cite{Ackland1997} 
to construct the dynamical matrix of the supercell.

\subsection{Energy-level renormalization}
\label{sec.si_results}

\begin{figure*}
\centering
\includegraphics[width=80mm]{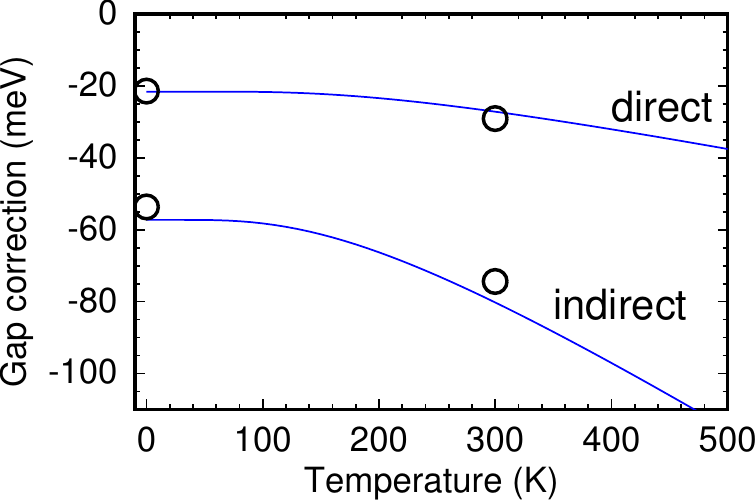}
\caption{
Calculated temperature dependent corrections to the direct and indirect gaps of silicon.
The blue lines were obtained over the entire temperature range from (\ref{eq.AH_shift}),
while the circles were calculated at specific temperatures (0~K and 300~K) using
(\ref{eq.sc-mehler-energy}).
\label{fig.si_energies} \vspace{-0.5cm}
} 
\end{figure*}

Silicon is an indirect gap semiconductor, with the valence band
maximum located at the $\Gamma$-point and the conduction band
minimum located close to the $X$-point.
With the ions frozen in their equilibrium positions we find 
a value of 2.55~eV for the direct gap at $\Gamma$ and a value
of 0.62~eV for the indirect $X$-$\Gamma$ gap.
We introduce the temperature-dependent corrections to
these gaps as  $(\<\e_\a\>_T - \<\e_\b\>_T)
- (\e_\a - \e_\b)$, where $\b$ refers to the occupied state
at the $\Gamma$ point and $\a$ refers to the unoccupied states
either at the $\Gamma$ point (direct gap) or $X$ point (indirect gap).

The values obtained for the gap corrections over the entire 
temperature range using (\ref{eq.AH_shift}) are shown as 
the blue lines in Fig.~\ref{fig.si_energies}.
We also plot as circles the corrections obtained at 0~K and 300~K
using (\ref{eq.sc-mehler-energy}).

In general, we see that the quantum motion of the nuclei acts
to close the band gap, albeit only by a small amount.
Using the quadratic expansion of (\ref{eq.AH_shift}), we obtain
zero-point corrections of -22 meV and -57~meV to the direct
and indirect gaps.
The zero-point correction obtained for the indirect gap
is close to the value of -52~meV found in recent calculations~\cite{Monserrat2014}.
At 300~K the magnitudes of these corrections increase slightly,
to 27 and 80~meV for the direct and indirect gaps respectively.

Using (\ref{eq.sc-mehler-energy}), we calculate corrections
of -21 and -54~meV to the direct and indirect gaps at 0~K,
and -29 and -74~meV at 300~K.
Within the error expected from our sampling procedure, these values
are equal to those obtained with (\ref{eq.AH_shift}).
Thus the neglect of the $\O(x_\nu^4)$ in the latter approach is
justified in this case.

\subsection{Phonon assisted absorption}

\begin{figure*}
\centering
\includegraphics[width=150mm]{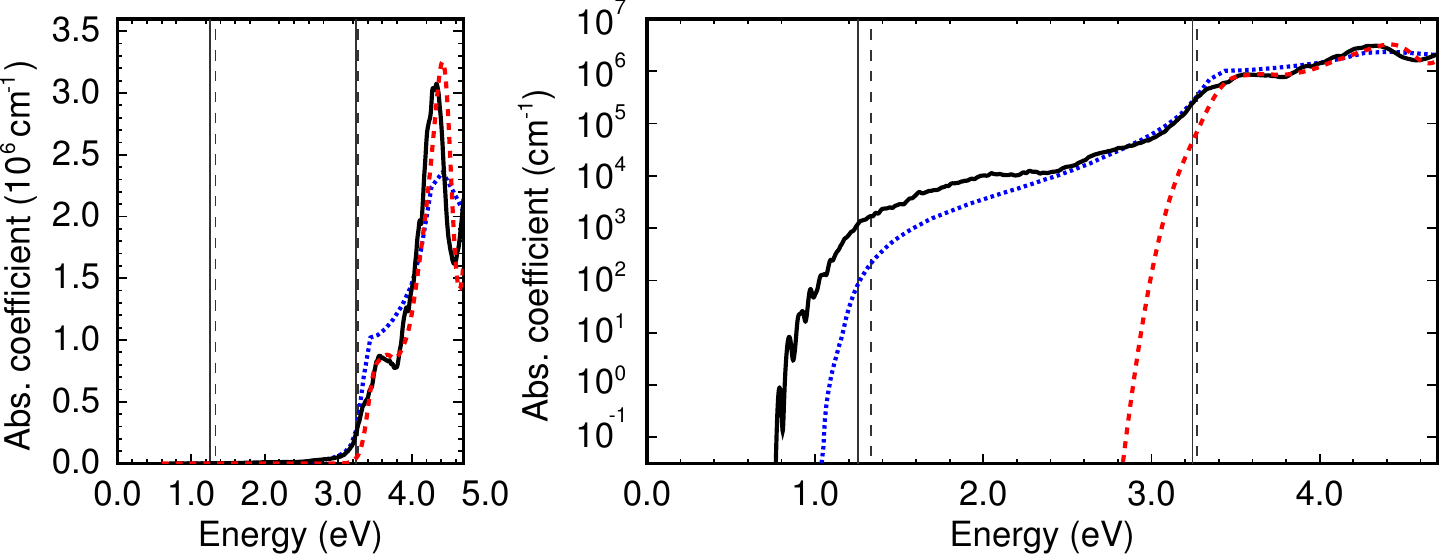}
\caption{
The absorption coefficient of bulk silicon, calculated with the ions fixed
in their equilibrium positions (red dashed lines) and with the semiclassical
expression (\ref{eq.sc-mehler-eps}) at 300~K (black line), compared to the 
experimental measurements of Ref.~\cite{Green1995} (blue dotted line).
The same data are shown on linear (left) and logarithmic (right) scales.
The vertical lines indicate the calculated band gaps (including scissor
correction), both with nuclei fixed in their equilibrium positions (dashed) and
after including temperature renormalization at 300~K, from Fig.~\ref{fig.si_energies} (solid).
\label{fig.si_abs} \vspace{-0.5cm}
} 
\end{figure*}

We now turn to the optical absorption spectrum.
In Fig.~\ref{fig.si_abs} we plot the absorption coefficient
obtained for the Si ions fixed in their equilibrium 
positions (red line), and compare to the experimental measurements
at 300~K reported in Ref.~\cite{Green1995} (blue line).
The theoretical fixed-ion spectrum displays
no absorption until the direct onset at 3.3~eV, while in experiment
indirect transitions are observed above the threshold of 1.1~eV.

When we evaluate
the semiclassical expression~(\ref{eq.sc-mehler-eps})
at 300~K (black line) we find that the theoretical
spectrum correctly displays absorption below the direct gap.
The absorption coefficient slowly increases in magnitude over the region 1.1--3.3~eV,
before a sharp increase above the direct gap threshold.

The semiclassical lineshape obtained here is not in perfect
agreement with experiment. We assign the discrepancy
to our use of a small supercell, equivalent to a $2\times2\times2$
$\vec{q}$-point sampling
(by contrast the calculations of Ref.~\cite{Noffsinger2012} used 
the Wannier interpolation scheme of Ref.~\cite{Giustino2007} to obtain extremely 
fine $\vec{q}$ grids, up to $40\times40\times40$).
The $2\times2\times2$ sampling is sufficient for us to at least observe the 
indirect $X-\Gamma$ transition,
but it is likely that the inclusion of more phonon modes will
redistribute the spectral weight in this energy region.
Investigating the convergence of the spectrum with increasing supercell size
is an important topic for future study.

Apart from this discrepancy, our calculated spectrum in Fig.~\ref{fig.si_abs}
illustrates that the semiclassical approximation discussed in section~\ref{sec.SC}
captures phonon-assisted indirect optical absorption in solids. Furthermore
this simple approach automatically incorporates the temperature dependence of the 
energy levels and their zero-point renormalization, shown by the vertical lines
in Fig.~\ref{fig.si_abs}.

\section{Choice of the electronic Hamiltonian}\label{sec.indir_sol_littleg}

Throughout this manuscript we employed a formalism based on the many-body electronic Hamiltonian
$H_e^R(r)$, wavefunctions $\Psi_\a^R(r)$, and eigenstates $E_\a^R$, see (\ref{eq.elecH}) 
and (\ref{eq.He}).
In practical calculations these quantities need to be replaced with appropriate approximations,
for example the Hartree-Fock method or the Kohn-Sham formulation of DFT~\cite{Kohn1965,Slater1951}.

If we consider Kohn-Sham DFT~\cite{Kohn1965}, then the electronic Hamiltonian at fixed nuclei
is replaced by the self-consistent Kohn-Sham Hamiltonian, and the many-body wavefunctions are replaced 
by Slater determinants of Kohn-Sham single-particle states, as we have done in
section~\ref{sec.Si}. After these substitutions all the formalism 
presented in this work remains essentially unchanged. For example, if the determinants 
$| \Psi_\a \>$ and $| \Psi_\b \>$ differ only in the occupations of the single-particle states 
$\psi_\a$ and $\psi_\b$, then the many-body electron-phonon matrix elements $G_{\a\b}^R$ introduced 
in (\ref{eq.bigG}) need to be replaced by matrix elements of of the self-consistent Kohn-Sham 
potential taken between these Kohn-Sham states, $g_{\a\b}^\nu = \< \psi_\a({\bf r}) | 
\Delta_\nu V_{\rm SCF}^R({\bf r})|\psi_\b({\bf r}) \>$ 
(with $\Delta_\nu = l_\nu \partial/\partial x_\nu$).
Similarly, the quadratic electron-phonon couplings 
$\partial \e_\a / \partial n_\nu$ in (\ref{eq.epcoupling})
need to be evaluated using differences in Kohn-Sham eigenvalues for the excitation energies.
Calculations of electron-phonon renormalization based on single-particle Hamiltonians (either empirical
or DFT) are abundant in the literature~\cite{Giustino20102,Marini2008, Capaz2005, Han2013,Gonze2011,
Monserrat2013,Ramirez2006,Patrick2013}, therefore it is expected that the formalism presented 
here will find immediate application in such calculations,
as demonstrated in the previous section.

In more sophisticated approaches it should be possible to describe excited states using the solutions
of the two-particle Bethe-Salpeter equation~\cite{Onida2002}: 
$|\Psi_\a\> = \sum_{cv} A_{cv}^\a a_c a^\dagger_v |\Psi_0\>$, with $a_c$ and $a_v^\dagger$ the operators 
for creating an electron and a hole in the single-particle states $\psi_c$ and $\psi_v$, respectively, 
and $A_{cv}^\a$ the Bethe-Salpeter eigenvector. In this case the many-body electron-phonon matrix element 
in (\ref{eq.bigG}) would incorporate both the eigenvectors $A_{cv}^\a$ and the Kohn-Sham electron-phonon
matrix elements $g_{cv}^\nu$. This alternative approach could be used to investigate exciton-phonon
interactions, provided practical approximations for the variation of the Bethe-Salpeter Kernel with 
the nuclear positions can be found, for example along the lines of Ref.~\cite{Beigi2003}.

The large flexibility afforded by the present formulation stems precisely from the choice
of introducing nuclear PES using (\ref{eq.PESe}), without making any assumptions on the underlying 
electronic Hamiltonian at fixed nuclei.

\section{Summary and conclusions}\label{sec.conclusions}

  \Table
  {\label{table}
  Summary of key results obtained in this work for quick reference. The first two sets
  apply to solids and molecules, respectively, and reflect the different choices of the
  non-interacting Hamiltonians. The last set provides a unified description of 
  electron-phonon effects in the optical spectra of solids and molecules, irrespective
  of system size.
  }
  \br
  &\hspace{-0.3cm}Solids&\\
  \mr
  &Description&Equation \\
  \\ \ns \ns \ns
  & Allen-Heine theory& (\ref{eq.AH_shift})\\
  & Zero-point renormalization& (\ref{eq.ZP})\\
  & No-phonon direct absorption & (\ref{eq.AH_trans_temp})\\
  & Phonon-assisted direct absorption & (\ref{eq.temp-dir-PA})\\
  & Phonon-assisted indirect absorption & (\ref{eq.ind-all2})\\
  & Exponential Urbach tail & (\ref{eq.my_urbach})\\
  \mr
  &\hspace{-0.3cm}Molecules&\\
  \mr
  &Description&Equation \\
  \\ \ns \ns \ns
  & Born-Huang expansion & (\ref{eq.BH}) \\
  & Franck-Condon theory & (\ref{eq.FC})& \\
  & Herzberg-Teller theory & (\ref{eq.HT})\\
  \mr
  &\hspace{-0.3cm}Molecules and Solids&\\
  \mr
  &Description&Equation \\
  \\ \ns \ns \ns
  & Semiclassical approximation to optical absorption & (\ref{eq.W-sc})\\
  & Temperature-dependent absorption using Mehler's formula & (\ref{eq.sc-mehler-T})\\
  \br
  \end{tabular}
  \end{indented}
  \end{table}

In this work we presented an attempt to place the theories of electron-phonon effects in 
the optical spectra of solids and molecules within a common framework.

We showed that the discussion of these phenomena in the quantum chemistry literature
and in the solid-state physics literature differ by the choice of the underlying
non-interacting electron-phonon Hamiltonian: in the case of molecules (which by extension
emcompasses the cases of point defects and Frenkel excitons in solids) the nuclei
experience a different potential energy surface for each electronic excitation,
whilst in the case of solids the nuclear dynamics is described by considering
only the PES generated by the electrons in their ground state. This subtle difference 
can be identified as the origin of the widely different approaches and methods developed 
for studying electron-phonon effects in chemistry and in physics.

Concentrating on molecules, we showed how well-established conceptual models
of electron-phonon effects in molecules, such as the Franck-Condon theory, the Born-Huang
expansion, and the Herzberg-Teller effect, can all be obtained by straightforward
low-order time-independent perturbation theory.

Along similar lines, we were able to derive the standard expression for indirect
optical absorption in solids using time-independent perturbation theory. In this
case our analysis revealed a number of subtle effects which have gone largely unnoticed
in the literature, for example we identified phonon-assisted optical absorption
in direct band gap materials.

The present work also allowed us to identify an exponential tail in the optical absorption
edge, which we tentatively assigned to the famous Urbach tail. To the best of our knowledge, 
this is the first time that an exponential edge emerges from a first-principles theory, 
while earlier proposals invariably used phenomenological models. 

We analyzed the formal basis of the Allen-Heine theory of temperature-dependent 
band structures. In this case we showed that the off-diagonal couplings between nuclear 
wavefunctions yield a correction to the zero-point renormalization not usually
considered in first-principles calculations.

Finally, we considered the semiclassical approach proposed in Ref.~\cite{Lax1952} for molecules
as an avenue to calculating optical absorption across the length scales. In particular we 
pointed out that the resulting expression avoids the difficulties associated with nuclear
wavefunctions corresponding to excited-state potential energy surfaces, 
and that it applies generally also to the case of solids.
We demonstrated an application of this expression by calculating
the phonon-assisted optical absorption spectrum of bulk silicon.
We provide a quick reference to our main results in Table~\ref{table}.

One important aspect of our theory is that the electron-phonon renormalization and the 
phonon-assisted optical absorption are described on the same footing. This strategy leads to a 
consistent theory of temperature-dependent optical absorption, and avoids the ambiguity that 
arises when trying to merge the theory of indirect absorption with that of temperature-dependent 
band structures.

In this work an effort was made to develop the theory by relying on a minimal set of approximations.
In order to keep the discussion accessible to the broadest audience we purposely refrained from 
making specific assumptions, e.g.\ the form of the electronic Hamiltonian at fixed nuclei or 
the translational invariance and the reciprocal space formalism for solids.
This choice should make it easier to tailor the present theory to specific applications, 
and work is currently in process to assess the performance of the formalism within 
the context of first-principles calculations.

It is hoped that the theory developed here will help clarifying the links between the many
different approaches to the electron-phonon problem, and will serve as a general and well defined
conceptual framework for future first-principles calculations of optical spectra.

\section*{Acknowledgements}
We thank E.\ Kioupakis and E. Yablonovitch for fruitful discussions, and
M. Ceriotti for bringing Sobol sequences to our attention.
This work was supported by the European Research Council (EU FP7 / ERC grant no. 239578),
the UK Engineering and Physical Sciences Research Council (Grant No. EP/J009857/1) and
the Leverhulme Trust (Grant RL-2012-001).

\appendix

\section{Normal modes of vibrations and ladder operators}
\label{app.phonons}

In order to make the manuscript self-contained we review the basic concepts and quantities
needed to describe the ground-state nuclear PES, $U^R_0$, in the harmonic approximation~\cite{Inksonbook}.
By expanding $U^R_0$ in powers of the nuclear displacements from their equilibrium geometry
$R_0$ and retaining terms up to second order (harmonic approximation) we have:
 \begin{equation}
  U^R_0 = U^{R_0}_0 + \frac{1}{2}\sum_{I \kappa, J \lambda} 
  \frac{\partial^2 U^R_0}{\partial R_{I \kappa} R_{J \lambda}} u_{I \kappa}u_{J \lambda},
 \end{equation}
where $u_{I \kappa}$ denotes the displacement from equilibrium of the $I$-th nucleus along
the Cartesian direction $\kappa$, and similary for $u_{J \lambda}$. From this expression
the dynamical matrix is introduced as:
  \begin{equation}
  D_{I \kappa, J \lambda} = \sqrt{\frac{1}{M_IM_J}} 
     \left. \frac{\partial^2U^R_0}{\partial R_{I \kappa} R_{J \lambda}} \right|_{R_0},
  \end{equation}
where $M_I$ and $M_J$ are the nuclear masses. Let us denote by  $e^\nu_{I \kappa}$ the eigenvector 
of this matrix for the eigenvalue $\Omega_{\nu}^2$. We can perform the transformation to normal 
mode coordinates $x_\nu$ as follows:
  \begin{equation}\label{eq.u-R}
  u_{I \kappa} = \sqrt{\frac{M_P}{M_I}} \sum_\nu e^\nu_{I \kappa} x_\nu,
  \end{equation}
where $M_P$ is a reference mass (usually the mass of a proton). In normal mode coordinates the PES becomes:
  \begin{equation}
  U_0^R = U^{R_0}_0 + {\sum}_\nu \frac{1}{2} M_P \Omega_{\nu}^2  x_{ \nu}^2.
  \end{equation}
By applying the same coordinate transformation to the kinetic energy we can rewrite (\ref{eq.nucH}) as:
  \begin{equation}\label{eq.harmonic}
  \sum_\nu\left[ - \frac{\hbar^2}{2M_P}\frac{\partial^2}{\partial x_\nu^2} 
    +  \frac{1}{2} M_P \Omega_{\nu}^2  x_{ \nu}^2 \right] \chi_{0 n}
   = \left[E_{0 n} -U^{R_0}_0 \right]\chi_{0 n}.
  \end{equation}
The solution of this equation is obtained as a product of independent quantum harmonic oscillators,
$\chi_{0 n}={\prod}_\nu \phi_{n_\nu}(x_\nu)$, with:
  \begin{equation}\label{eq.phi}
  \phi_{n_\nu}(x_\nu) = \frac{(2\pi l_\nu^2)^{-1/4}}{\sqrt{2^{n_\nu} n_\nu!}} 
  \exp\left(-\frac{x_\nu^2}{4 l_\nu^2}\right)H_{n_\nu}\left(\frac{x_\nu}{\sqrt{2}l_\nu}\right),
  \end{equation}
and energy $E_{0 n_\nu} = \hbar\Omega_\nu ( 1/2 + n_\nu)$. In this equation $H_{n_\nu}(x)$ is 
the Hermite polynomial of order $n_\nu$ and $l_\nu$ is defined as in (\ref{eq.epcoupling2}). 
The set of quantum numbers $n_1, n_2,\dots$ defines the composite index $n$ in the wavefunction 
$\chi_{0 n}$, and identify the occupations of each vibrational quantum state. It is customary 
to indroduce the ladder operators $b_\nu^\dagger$ and $b_\nu$ such that:
  \begin{eqnarray}
  b^\dagger_\nu |\phi_{n_\nu}\> = \sqrt{n_\nu+1}|\phi_{n_\nu + 1}\>,  \qquad
  b_\nu |\phi_{n_\nu}\> = \sqrt{n_\nu}|\phi_{n_\nu - 1}\>. \label{eq.app.1}
  \end{eqnarray}
These operators have the following useful properties which are used repeatedly throughout the
manuscript:
  \begin{eqnarray}
  [b_\nu,b^\dagger_{\mu}] = \delta_{\nu\mu}, \\
  b^\dagger_\nu b_\nu |\phi_{n_\nu}\> = n_\nu |\phi_{n_\nu}\>, \\
  x_\nu = l_\nu (b^\dagger_\nu + b_\nu), \label{eq.x-b} \\
  x_\nu^2 =l_\nu^2(b^\dagger_\nu b^\dagger_\nu + b_\nu b_\nu + 2b^\dagger_\nu b_\nu + 1). \label{eq.app.2}
  \end{eqnarray}
In addition the kinetic energy operator can be rewritten in terms of ladder operators as:
  \begin{eqnarray}
  T^R = -\frac{1}{4}\sum_\nu \hbar\Omega_\nu (b^\dagger_\nu - b_\nu)^2.
  \end{eqnarray}
Using (\ref{eq.app.1}) and (\ref{eq.app.2}) the expectation value of the square displacement is obtained as:
  \begin{eqnarray}\label{eq.xnu2}
  \<\phi_{n_\nu}| x_\nu^2 |\phi_{n_\nu}\> = l_\nu^2 (2 n_\nu + 1),
  \end{eqnarray}
and the corresponding thermal average is given by:
  \begin{eqnarray}\label{eq.x2-T}
  \<x^2_\nu\>_T = \frac{1}{Z}{\sum}_{n_\nu=0}^\infty \exp\left[-\frac{E_{0 n_\nu}}{\kT} \right]
  l_\nu^2(2 n_\nu + 1) = l_\nu^2 [2 n_B(\Omega_\nu,T) + 1],
  \end{eqnarray}
with $Z = {\sum}_{n_\nu=0}^\infty \exp(-E_{0 n_\nu}/\kT)$ being the canonical partition function 
and $n_{\rm B}$ the Bose-Einstein distribution. The total energy of the state $E_{0 n}$ is:
  \begin{equation}\label{eq.sho_en}
  E_{0 n}  = U^{R_0}_0 + {\sum}_\nu \hbar\Omega_\nu\left[ \frac{1}{2} + n_\nu\right].
  \end{equation}
The eigenstates of the quantum harmonic oscillator have the following useful property:
  \begin{equation}\label{eq.mehler}
   \frac{1}{Z}{\sum}_{n_\nu=0}^\infty e^{-\frac{E_{0 n_\nu}}{\kT}}
   |\phi_{n_\nu}(x_\nu)|^2 = \frac{1}{\sqrt{2\pi \<x^2_\nu\>_T}}\exp\left[-\frac{x^2}{2 \<x^2_\nu\>_T } \right],
  \end{equation}
with $\<x^2_\nu\>_T$ given by (\ref{eq.xnu2}). This result derives directly from
(\ref{eq.phi}) and Mehler's formula~\cite{Watson1933}:
  \begin{eqnarray}
  \sum_{n=0}^\infty \frac{\exp\left[-(x^2+y^2)/2\right]}{2^n n! \sqrt{\pi}} t^n H_n(x)H_n(y) = \nonumber \\
    \qquad \qquad \frac{1}{\sqrt{\pi(1-t^2)}}\exp\left[\frac{4xyt - (x^2+y^2)(1+t^2)}{2(1-t^2)} \right],
  \end{eqnarray}
with $x$, $y$, and $t$ real numbers.

\section{Perturbative expansions}

\subsection{General formulas}\label{app.pert-exp}

All the results presented in this work are based on standard time-independent nondegenerate
perturbation theory~\cite{Sakuraibook}. For ease of reference we report here the key equations
employed throughout the manuscript. The eigenstates of total Hamiltonian of the joint electron-nuclear 
system, $\H$ in (\ref{eq.fullH}), are denoted by $|\a n^{\rm e}\>$, with the superscript standing for
``exact''. The corresponding energy is denoted as $E_{\a n}^{\rm e}$: $\H |\a n^{\rm e}\> = 
E_{\a n}^{\rm e} |\a n^{\rm e}\>$.  We express the total Hamiltonian as the sum of a ``non-interacting'' 
Hamiltonian, $H_0$, and a perturbation, $\Delta H$, so that the non-interacting eigenstates and eigenvalues
are given by: $H_0 |\a n\> = E_{\a n} |\a n\>$. The second-order perturbative expansion of the 
energy $E_{\a n}^{\rm e}$ is given by:
  \begin{eqnarray}
  E_{\a n}^{\rm e} = E_{\a n} + \<\a n| \Delta H | \a n\>
  + {\sum_{\b m}}' \frac{|\<\b m| \Delta H | \a n\>|^2}{E_{\a n} - E_{\b m}} + \O(3),
  \label{eq.en_exp} 
  \end{eqnarray}
where the prime on the summation is to exclude the term with $\a=\b$ and $n = m$.
The first-order expansion of the eigenstate $|\a n^{\rm e}\>$ is:
  \begin{eqnarray}
  |\a n^{\rm e}\> &=& |\a n\> + {\sum_{\b m}} ' \frac{\<\b m| \Delta H | \a n\>}{E_{\a n} - E_{\b m}} |\b m\>
  + \O(2). \label{eq.stat_exp}
  \end{eqnarray}
In both expressions the notation $\O(n)$ stands to indicate that the neglected terms are
proportional to $\<\b m| \Delta H | \a n\>^n$.

The choice of the non-interacting Hamiltonians in section~\ref{sec.pert} guarantees
that the use of perturbation theory is legitimate. In fact, if we consider for instance the case 
of solids, the size of the first order correction to the energy in (\ref{eq.en_exp}) is of the order 
of the ratio between a characteristic phonon energy and the band gap, 
$\hbar\Omega_\nu/E_{\rm g}$~\cite{Cardona2005}.

\subsection{Degeneracies in the electronic spectrum}\label{app.degen}

In this work we considered exclusively the case of non-degenerate perturbation theory.
In the case that electronic states calculated for nuclei in their equilibrium positions $R_0$
are degenerate, i.e.\ $E_\a = E_\b$, there exists an ambiguity in the perturbation
expansion. This is best seen by considering (\ref{eq.elec_en}), which shows that the
energy denominator $E_\a-E_\b$ yields a singularity. In this case it is necessary to
repeat the entire set of derivations presented in this work using {\it degenerate}
perturbation theory~\cite{Sakuraibook}. Following the prescription of degenerate perturbation 
theory, we would need to set the gauge of the wavefunctions in the degenerate subspace
by diagonalizing the perturbation $\Delta H$ in the same subspace. 

The treatment of electronic degeneracies does not pose any problems in the semiclassical
approach described in section~\ref{sec.SC}, since the degeneracy is traced out in the
evaluation of the optical absorption spectra. This is easily understood by noting that
the key equation (\ref{eq.W-sc}) does not contain any energy denominators.
The situation is more complex in the solid-state picture and in the molecular picture
described in sections~\ref{sec.hamiltonians} and \ref{sec.ham-solids}, respectively.
In fact in the molecular case electronic degeneracies correspond to crossings of 
potential energy surfaces. These crossings are responsible for non-adiabatic couplings
and have been the subject of numerous investigations in the quantum chemistry 
literature~\cite{Domcke2012,Bersukerbook}. In the case of solids the 
electronic degeneracies can be addressed by modifying the definition of the perturbative 
correction in (\ref{eq.AH}) in such a way as to treat the non-degenerate and the degenerate 
parts separately.

\subsection{Order of perturbative corrections}\label{app.rcube}

Throughout the manuscript we indicated the order of the perturbative expansion using alternatively
the notation $\O(n)$ or $\O(x_\nu^n)$. This distinction is important in the study of electron-phonon
interactions, as already pointed out in Ref.~\cite{Allen1976}.

In order to make this point clear we consider the matrix elements for optical transitions 
in solids, $\<\b m^{\rm e,s}|\Delta|\a n^{\rm e,s}\>$. In (\ref{eq.wholerate}) this matrix element
is expanded up to order $\O(2)$ in the perturbation $H_e^R-H_e^{R_0}$. However, the terms
$\e_\a^R$ and $V_{\a\b}^R$ appearing in (\ref{eq.dir-PA})--(\ref{eq.ind-PA}) can be expanded further
in terms of nuclear displacements to arbitrary order. For example, these terms are given in
(\ref{eq.V-x})--(\ref{eq.epsa-x}) up to an error of order $\O(x_\nu^2)$.

More generally, an expansion to order $\O(n)$ always implies that the expression is also correct
up to errors of order at least $\O(x_\nu^n)$, but the reverse is not true in general.
For this reason it is important to always follow the sequence of expanding first in terms
of the Hamiltonian perturbation $\Delta H$, and then in terms of the nuclear displacements.

While this simple rule appears obvious, a considerable amount of debate in the literature
was generated precisely by confusion on this point, most notably the link between the Fan and
Debye-Waller corrections to the band structures of solids~\cite{Allen1976}.

\section*{References}
\bibliography{papers}

\end{document}